\documentclass[aps,prb,reprint, superscriptaddress]{revtex4-2}
\usepackage[english]{babel}
\usepackage[utf8]{inputenc}
\usepackage{amsmath}
\usepackage{amsfonts}
\usepackage{amssymb}
\usepackage{mathtools}
\usepackage{braket}
\usepackage{graphicx}
\usepackage{grffile}
\usepackage{siunitx}
\usepackage{rotating}
\usepackage{bbm}
\usepackage{multirow}
\renewcommand{\d}{\mathrm{d}}
\newcommand{\e}{\mathrm{e}}

\begin{document}
	
	\title{Population Control Bias and Importance Sampling in Full Configuration Interaction Quantum Monte Carlo} 
	\author{Khaldoon Ghanem}
	\affiliation{%
		Max Planck Institute for Solid State Research, Heisenbergstr.\ 1, 70569 Stuttgart, Germany%
	}
	\email{k.ghanem@fkf.mpg.de}
	
	\author{Niklas Liebermann}
	\affiliation{%
		Max Planck Institute for Solid State Research, Heisenbergstr.\ 1, 70569 Stuttgart, Germany%
	}
	
	\author{Ali Alavi}
	\affiliation{%
		Max Planck Institute for Solid State Research, Heisenbergstr.\ 1, 70569 Stuttgart, Germany%
	}
	\affiliation{%
		Dept of Chemistry, University of Cambridge, Lensfield Road, Cambridge CB2 1EW, United Kingdom%
	}%
	\email{a.alavi@fkf.mpg.de}
	
	\date{\today}

	\begin{abstract}
		Population control is an essential component of any projector Monte Carlo algorithm.
		This control mechanism usually introduces a bias in the sampled quantities that is inversely proportional to the population size.
		In this paper, we investigate the population control bias in the full configuration interaction quantum Monte Carlo method. 
		We identify the precise origin of this bias and quantify it in general.
		We show that it has different effects on different estimators and that the shift estimator is particularly susceptible.
		We derive a re-weighting technique, similar to the one used in diffusion Monte Carlo, for correcting this bias and apply it to the shift estimator.
		We also show that by using importance sampling, the bias can be reduced substantially.
		We demonstrate the necessity and the effectiveness of applying these techniques for sign-problem-free systems where this bias is especially notable.
		Specifically, we show results for large one-dimensional Hubbard models and the two-dimensional Heisenberg model, where corrected FCIQMC results are comparable to the other high-accuracy results.
	\end{abstract}
	\maketitle
	
	\section{Introduction}
	Projector Monte Carlo algorithms, including Green's function Monte Carlo (GFMC)~\cite{Lee1992}, diffusion Monte Carlo (DMC)~\cite{Reynolds1990},  and full configuration interaction quantum Monte Carlo (FCIQMC)~\cite{Booth2009}, have become indispensable tools in extracting the ground state properties of various quantum systems.
	These methods employ a stochastic version of the power method; a method that starts from an initial wavefunction with a non-zero overlap with the ground state and filters out higher excited states by a repeated application of a suitable imaginary-time propagator of the Schr\"odinger equation.
	As the system size grows, its Hilbert space grows exponentially, and a stochastic representation of the wavefunction becomes a necessity.
	The wavefunction is then sparsely represented by a set of walkers that are evolved according to the propagator.
	Their expectation value on any configuration reflects the wavefunction value on that configuration.
	
	The main challenge faced by these algorithms, compared to traditional stochastic methods, is that their propagators do not form column-stochastic matrices in general~\cite{Hetherington1984}.
	On the one hand, their elements can be either positive or negative.
	On the other hand, the sum of the absolute values for each column is not normalized.	
	The first issue implies that walkers should be signed and often leads to a sign problem. 
	The latter issue can be dealt with using a branching process where walkers need to be added or removed dynamically to reflect the changes in the $L_1$-norm of the wavefunction.
	However, the branching creates a practical problem because there is a chance that all walkers will die or their storage exceeds the available computer memory.
	Moreover, the computational effort per sample and the statistical accuracy are directly proportional to the total number of walkers, so minimizing the fluctuations in the number of walkers makes running times more predictable and markedly enhances the utilization of available computational resources.
	Consequently, different projector Monte Carlo methods apply some  \emph{population control} mechanism that keeps the number of walkers around a set target throughout the simulation.
	It has long been known that this population control introduces a bias in DMC calculations~\cite{Umrigar1993} and other Green's function methods~\cite{Ceperley1986} and that the bias scales inversely with the number of walkers~\cite{Hetherington1984}.
	Different remedies have been proposed to overcome this bias, including an extrapolation in the number of walkers, a modification of the wavefunction renormalization procedure, and deriving new unbiased estimators by reweighting with the discarded scale factors~\cite{Cerf1995}.	
	
	The population control bias has often been neglected in FCIQMC because it is usually much smaller than the statistical error bars or other biases~\cite{Vigor2015}.
	Indeed, for any system with a fermionic sign problem, the original full FCIQMC algorithm needs a minimum number of walkers, below which the sign-incoherent noise dominates the simulation and the dynamics of the walkers become unstable.
	This number of walkers is typically large enough that the population bias does not exceed the statistical error bars.
	Moreover, in practice, the initiator approximation is often used instead of the original FCIQMC algorithm.
	This approximation constrains the dynamics of the walkers, which prevents the sign-incoherent noise from propagating and destabilizing the simulation~\cite{Cleland2010}.
	The constraint allows stable simulations using any number of walkers but introduces a bias that also scales down with the number of walkers.
	The resulting initiator bias is much larger than the population control bias, and reducing it requires increasing the population.
	Therefore, the population bias has been masked in these applications by the initiator bias and did not pose a practical concern.
	In contrast to these systems, there are some sign-problem-free ones, where no minimum number of walkers is required for getting stable FCIQMC simulation, and it is a matter of obtaining more samples to reduce the statistical error bars.
	For such systems, the population control bias is the only systematic bias, making its study most convenient in these cases.
	Consequently, in this paper, we study the population control bias in FCIQMC, particularly in its application to sign-problem-free systems, and in conjunction with importance sampling.
	
	The paper is structured as follows:
	We first review the FCIQMC method, its sign problem, energy estimators, and its population control schemes.
	We show that population control leads to a biasing source term in the master equation, which equals the covariance between the shift parameter and the wavefunction.
	We also show that this covariance term scales inversely with the number of walkers and use it to relate the biases of different energy estimators.
	Ref.~\cite{Vigor2015} borrowed a reweighting procedure developed originally for DMC and showed numerically that it effectively corrects the projected energy estimates.
	We provide a derivation of this reweighting procedure within FCIQMC and adapt it for correcting the shift estimator.
	We then examine the effect of the importance sampling on reducing the bias.
	Finally, we demonstrate the effectiveness of the correction technique and the importance sampling by applying them to the one-dimensional Hubbard model and the two-dimensional Heisenberg model with large lattice sizes.
	
	\section{Background}
	\subsection{FCIQMC}
	FCIQMC projects  the ground state of a Hamiltonian $\hat{H}$ from an initial state $\ket{\psi_0}$  by evolving it according to the imaginary-time Schr\"odinger equation
	\begin{equation}\label{eq:schrd}
	- \frac{\d}{\d \tau} \ket{\psi(\tau)} = \bigl[ \hat{H}-S\bigr] \ket{\psi(\tau)}
	\end{equation}
	where $S$  is some scalar shift of the diagonal elements.
	The formal solution of this equation can be written in terms of the eigenstates of  the Hamiltonian $\ket{\phi_n}$ and their energies $E_n$ as following
	\begin{align}
	\ket{\psi(\tau)} &= \e^{-(\hat{H}-S) \tau} \ket{\psi_0} = \sum_n \braket{\phi_n|\psi_0}  \e^{-(E_n-S) \tau} \ket{\phi_n} \;.
	\end{align}
	If the shift $S$ is lower than the first excited state energy, then contributions from all excited states decay exponentially leaving only the ground state wavefunction in the long time limit
	\begin{equation}
	\label{eq:converge}
	\lim_{\tau \to \infty} \ket{\psi(\tau)} = \braket{\phi_0|\psi_0} \e^{-(E_0-S)\tau} \ket{\phi_0} \;.
	\end{equation}
	
	The wavefunction in  FCIQMC is represented by a set of walkers that occupy the Hilbert space spanned by Slater determinants, which are constructed from a finite basis of single-particle orbitals. 
	The walker population is updated stochastically at each time step  such its average change respects the time-discretized Schr\"odinger equation  i.e.
	\begin{equation}\label{eq:master}
	-{\frac{\Delta \overline{N_i}(\tau) }{\Delta \tau}} = \bigl[H_{ii} - S\bigr] N_i(\tau) + \sum_{i\neq j} H_{ij} N_j(\tau)\;, 
	\end{equation}
	where $\Delta\tau$ is the discretized time-step, $N_i \coloneqq \braket{D_i|\psi}$ is the number of walkers on determinant $\ket{D_i}$, and $H_{ij} \coloneqq \braket{D_i|\hat{H}|D_j}$ are the Hamiltonian matrix elements in this determinant basis.
	In the long time limit, the ground state wavefunction can then be estimated by averaging the walkers over many time steps
	\begin{equation}
	\overline{N_i} \approx \braket{D_i|\phi_0}  \;.
	\end{equation}
	
	The master equation \eqref{eq:master} is implemented using three steps of walker dynamics~\cite{Booth2009}:
	\begin{itemize}
		\item \emph{Spawning:} Each walker on determinant~$\ket{D_i}$ randomly spawns another walker on a connected determinant~$\ket{D_j}$ with probability~$P_{i\rightarrow j}$ .
		The spawned walker is then accepted with probability
		\begin{equation}
		P_\text{spawn} = \Delta \tau \frac{\lvert H_{ij} \rvert}{P_{i\rightarrow j}}\;,
		\end{equation}
		and it is given a sign opposite to the sign of $H_{ij} N_i$.
		When $P_\text{spawn}>1$, one spawn is created stochastically with a probability that equals the fraction part, while the integral part determines the number of additional deterministically-created spawns.
		\item \emph{Death/Clone: }
		Each walker dies or gets cloned with a probability that equals the absolute value of
		\begin{equation}
		P_\text{death/clone} = \Delta \tau \bigl(H_{ii} - S\bigr)\;.
		\end{equation}
		The walker is killed when this value is positive and duplicated when it is negative.
		\item \emph{Annihilation: }
		Walkers of opposite sign residing on the same determinant are cancelled and removed from the simulation.
		As discussed next, this step is crucial to ensure convergence to the ground state.
		Without it, the solution would quickly get contaminated by noise that masks the physical ground state.
	\end{itemize}
	
	\subsection{Sign Problem}
	In the absence of annihilation, the in-phase combination of the positive and negative walkers corresponds to a non-physical solution that grows faster than the desired out-of-phase solution~\cite{Spencer2012}.
	The annihilation reduces the growth rate of the in-phase solution, and the reduction is proportional to the total number of walkers. 
	As a result, a minimum number of walkers, known as the annihilation plateau, is needed in order for the physical solution to outgrow the in-phase one and for the correct sign structure to emerge.
	This minimum number is typically less than the full size of the Hilbert space but can be a substantial fraction of it.
	
	The initiator approximation overcomes this issue by modifying the dynamics of the walkers~\cite{Cleland2010}.
	Only walkers residing on determinants with a population above a certain threshold are allowed to spawn onto empty determinants.
	This prevents lowly-populated determinants (the non-initiators) with fluctuating signs from propagating the sign-incoherent signal further.
	This constrained dynamics allows FCIQMC to converge at an arbitrarily small number of walkers.
	The cost is introducing a bias that can be systematically improved by increasing the number of walkers. 
	The initiator bias can also be significantly reduced using the recently-proposed adaptive shift method~\cite{Ghanem2019, Ghanem2020}. 
	In this method, the life time of a non-initiator determinant is boosted by reducing its shift to account for the missing back-spawns from its underpopulated surrounding.
	
	The annihilation plateau is the manifestation of the sign problem in FCIQMC.
	There are a few model systems, however, which are free from this sign problem and can converge at any number of walkers without the initiator constraint.
	For these systems, all walkers spawned onto a specific determinant have the same sign.
	One obvious way this can happen is when all the non-diagonal elements of the Hamiltonian have a negative sign (so-called stoquasticity); thus, all walkers are of the same sign.
	More generally, a Hamiltonian is sign-problem-free when we can classify the determinants into two disjoint sets such that the matrix elements between members of the same set are negative (or zero), while the matrix elements between members of different sets are positive (or zero).
	In this case, the walkers of the two sets have opposite signs, but they never mix throughout the simulation.
	This general case can always be recast as the earlier one-sign case by using a different sign convention for the determinants in one of the two sets, which would make all the non-diagonal matrix elements negative.  
	Examples of sign-problem-free systems include the one-dimensional Hubbard model with open boundary conditions, the one-dimensional Hubbard model with periodic boundary conditions and specific combinations of up- and down-electrons, and the two-dimensional Heisenberg model on a bipartite lattice. All of these systems are introduced and discussed in Section~\ref{sec:applications}.
	
	\subsection{Projected Energy}
	Two kinds of energy estimators are used in FCIQMC.
	One is the energy shift discussed in the next section.
	Another is the so-called projected energy obtained by projecting on some trial wavefunction~$\ket{\psi^\mathrm{T}}$:
	\begin{equation}
	E^\mathrm{T} \coloneqq \frac{\braket{\psi^\mathrm{T}|\hat{H}|\psi}}{\braket{\psi^\mathrm{T}|\psi}}\;.
	\end{equation}
	As the average wavefunction converges to the true ground state, its projected energy converges to true ground-state energy.
	To compute this estimate,  one does not need to know the average wavefunction explicitly.
	Instead, the numerator and denominator are computed out of the instantaneous wavefunctions during the simulation and then averaged separately.
	The ratio of the averaged quantities gives an estimate of the projected energy of the average wavefunction:
	\begin{equation}\label{eq:pe_estimate}
	E^\mathrm{T} \approx \frac{\sum_{i=1}^L  \braket{\psi^\mathrm{T}|\hat{H}|\psi(\tau_i)}}{\sum_{i=1}^L \braket{\psi^\mathrm{T}|\psi(\tau_i)}}\;,
	\end{equation}
	where $L$ is the number of FCIQMC samples~$\ket{\psi(\tau_i)}$ taken after an appropriate thermalization period.
	
	For single-reference molecular systems, the trial wavefunction is typically chosen to be the reference determinant with the highest population (usually the Hartree--Fock determinant).
	For multi-reference molecular systems, the simulation is run for some period before a trial space is constructed out of the most populated determinants.
	The ground state in the trial space is then calculated and used as a trial wavefunction for a more reliable projected energy estimator. 
	However, for some model systems, like the ones considered in this paper, the wavefunction spreads across the whole Hilbert space such that the overlap with a restricted subspace is very small, and the energy estimates mentioned above are very noisy.
	For these systems, we introduce a particular projected energy estimator, which is useful when they are free of the sign problem.
	Let us denote by $\ket{\pm \mathbbm{1}}$, the uniform trial wavefunction that spans the whole Hilbert space equally and has the same sign structure as the ground state wavefunction
	\begin{equation}\label{eq:uniform_trial}
	\braket{D_i|{\pm \mathbbm{1}}} \coloneqq  \mathcal{S}_i  = \begin{cases}
	+1 \text{ when } \braket{D_i|\phi_0} >0\\
	-1 \text{ when } \braket{D_i|\phi_0} <0\\
	\end{cases}\;,
	\end{equation}
	the corresponding projected energy estimator then reads
	\begin{equation}\label{eq:uniform_pe}
	E^{\pm 1} \coloneqq \frac{\braket{\pm \mathbbm{1}|H|\psi}}{\braket{\pm  \mathbbm{1}|\psi}} = \frac{\sum_i{|\overline{N_i}| \sum_j \mathcal{S}_i \mathcal{S}_j  H_{ij}}}{\sum_i{|\overline{N_i}|}}\;.
	\end{equation}
	We call this the \emph{uniform projected energy.}
	Calculating this estimator requires knowing the sign structure of the ground state a priori, which is feasible in sign-problem-free cases.
	Assuming in these cases a sign convention that renders all walkers positive, the uniform projected energy takes the following simple form
	\begin{equation}
	E^{\pm 1} = \frac{\sum_i{\overline{N_i} \sum_j H_{ij}}}{\overline{N}}\;,
	\end{equation}
	where $N$ is the total number of walkers.
	This estimator will be relevant later when deriving a correction of the population control bias.
	
	\subsection{Population Control}
	From Eq.~\eqref{eq:converge}, it is clear that unless the shift~$S$ equals the ground state energy~$E_0$, the overall amplitude of the wavefunction, and thus the total number of walkers~$N$, would decay/grow exponentially. 
	To maintain a stable number of walkers in the long-run, the shift is updated periodically as following:
	\begin{equation}\label{eq:control}
	S(\tau) = S(\tau-A\Delta\tau) - \frac{\gamma}{A\Delta\tau} \ln \frac{N(\tau)}{N(\tau-A\Delta\tau)}\;,
	\end{equation}
	where $\gamma$ is a damping factor, and $A$ is the number of time steps between successive updates of the shift. The intuition behind this formula is simple. When the population increases, the shift is lowered, which decreases the population in subsequent steps and vice versa. 
	The logarithm is used because the change in the number of walkers is exponentially proportional to the energy/shift difference. 
	
	This population control scheme is the most commonly used one in FCIQMC.
	To get the number of walkers around a set target, the shift is held constant at the beginning of the simulation, during which the number of walkers increases exponentially. Then once it reaches its desired value,  it is stabilized using Eq.~\eqref{eq:control}.
	After an equilibration period, the average number of walkers becomes a constant, and thus the average value of the shift can be used as an independent energy estimator.
	
	Ref.~\cite{Yang2020} has recently proposed an improved scheme that avoids potential overshoots in the number of walkers. 
	The new scheme actively targets the desired number of walkers in its update formula, evading the need for the initial constant-shift mode.
	Other population control schemes can also be devised where the amplitude of the wavefunction is controlled using its overlap with a trial wavefunction $\braket{\psi^\mathrm{T}|\psi}$ instead of the total number of walkers~$N$.
	This can be achieved either using a formula similar to Eq.~\eqref{eq:control} or by setting the shift equal to the corresponding projected energy estimator $E^\mathrm{T}$ at each time step.
	The latter keeps the average value of the overlap $\braket{\psi^\mathrm{T}|\psi}$ constant because
	\begin{equation}
	- \frac{\d}{\d \tau} \langle\psi^\mathrm{T}\overline{\ket{\psi(\tau)}} = \bigl[E^\mathrm{T}(\tau) - S(\tau)\bigr] \braket{\psi^\mathrm{T}|\psi(\tau)}\;.
	\end{equation}
	Such intermediate normalization schemes have the advantage of reducing the statistical noise in the corresponding projected energy estimates.
	
	\section{Population Control Bias}

	Population control introduces a bias into the stochastic solution of FCIQMC  because of the dependence of the shift on the wavefunction itself and the resulting feedback into the solution.
	In the following, we will spell out this dependence explicitly and gain insights into the associated bias.
	
	At each time step of FCIQMC, the wavefunction is updated stochastically such that the average change in the wavefunction obeys the Schr\"odinger equation~\eqref{eq:schrd} with shift~$S(\tau)$
	\begin{equation}\label{eq:inst_master}
	- \frac{\d}{\d\tau} \overline{\ket{\psi(\tau)}} = \bigl[ \hat{H}-S(\tau)\bigr] \ket{\psi(\tau)}\;.
	\end{equation}
	The master equation of the \emph{average} wavefunction itself, however, is obtained by taking the average of both sides~\footnote{
		The average in Eq.~\eqref{eq:inst_master} is conditional on the instantaneous wavefunction and shift values from the previous iteration, while the average in  Eq.~\eqref{eq:avg_master} is the expectation value over all the possible histories of the simulation.
		We denote both these averages by an overline, sacrificing some mathematical rigor for more readability.
		The intended meaning should be clear from the context.
	}
	\begin{equation}\label{eq:avg_master}
	- \frac{\d}{\d\tau} \overline{\ket{\psi(\tau)}} =  \hat{H} \overline{\ket{\psi(\tau)}} - \overline{S(\tau) \ket{\psi(\tau)}}\;.
	\end{equation}
	Controlling the population by updating the shift necessarily implies that the shift and the wavefunction are correlated, and thus the average product 
	$\overline{S(\tau) \ket{\psi(\tau)}}$
	cannot be replaced by the product of the averages, but there is a non-vanishing covariance term
	\begin{equation}
	\overline{S(\tau) \ket{\psi(\tau)}} = \overline{S(\tau)} \ \overline{\ket{\psi(\tau)}} + \operatorname{Cov}[S(\tau), \psi(\tau)]\;.
	\end{equation} 
	As a result, the average wavefunction does not satisfy the  Schr\"odinger equation exactly, but it instead evolves according to the following equation, which has an additional source term
	\begin{equation}\label{eq:master_biased}
	- \frac{\d}{\d\tau} \overline{\ket{\psi(\tau)}} = \bigl[ \hat{H}-\overline{S(\tau)}\bigr] \overline{\ket{\psi(\tau)}} -  \operatorname{Cov}[S(\tau), \psi(\tau)]\;.
	\end{equation}
	This additional term is the origin of the population control bias in FCIQMC.
	
	We can go one step further and express the above equation in terms of the average walker populations
	\begin{align}\label{eq:master_biased2}
	-{\frac{\d}{\d \tau}} \overline{N_i}= \left[H_{ii} + \frac{B_i}{\overline{N}} - \overline{S}\right] \overline{ N_i} + \sum_{j\ne i} H_{ij} \overline{N_i} \;,
	\end{align}
	where we defined the relative bias of population $N_i$ as
	\begin{equation}\label{eq:rel_bias}
	B_i \coloneqq {-\operatorname{Cov}}\left(N_i, S\right) \frac{\overline{N}}{\overline{N_i}} \;.
	\end{equation}
	From this, one can view the population bias as a modification of the diagonal elements of the Hamiltonian
	\begin{equation}\label{eq:ham_bias}
	H^\prime_{ii}= H_{ii}  +   \frac{B_i}{\overline{N}} \;,
	\end{equation}
	and so the biased wavefunction as the ground state of this modified Hamiltonian rather than the desired original Hamiltonian.
	Note that in anticipation of the discussion next subsection, we defined $B_i$ in a way that explicitly highlights the dependence of the bias on the average number of walkers $\overline{N}$.

	\begin{figure}
		\center
		\includegraphics[width=\columnwidth]{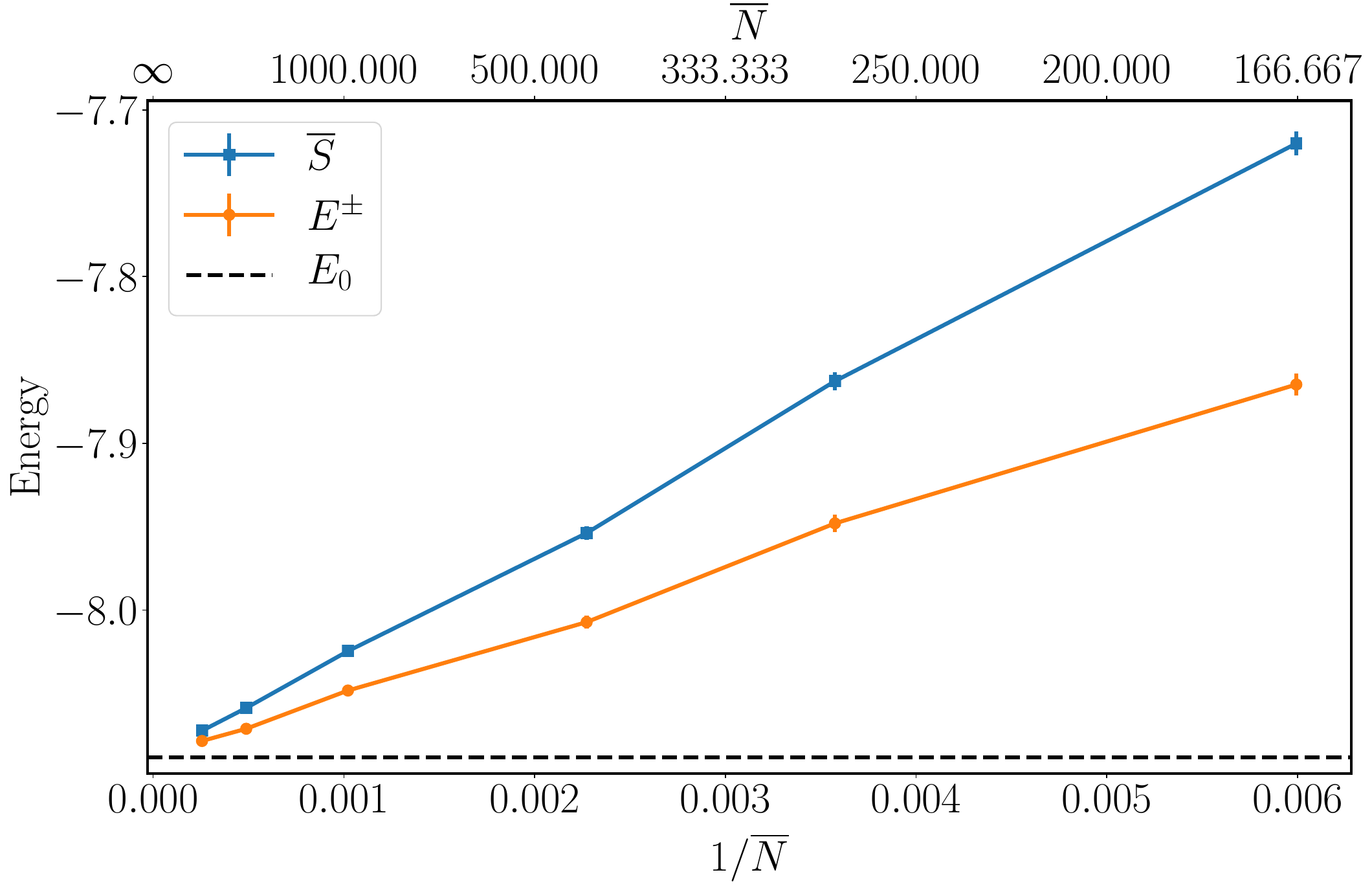}
		\caption{\label{fig:scaling}
			FCIQMC results for the 14-site one-dimensional Hubbard model for a different number of walkers.
			The projected energy $E^\pm$ is computed using a uniform trial wavefunction as defined by Eq.~\eqref{eq:uniform_trial}.
			The set target numbers of walkers are $100, 200, ...,3200$, while their corresponding average number of walkers $\overline{N}$ are always larger due to the overshoot in the constant-shift mode.
		}
	\end{figure}
	As an illustrative example, we show in Fig.~\ref{fig:scaling} the energy estimates for the one-dimensional 14-site Hubbard model with $U/t = 4$ and periodic boundary conditions.
	The model is small enough that we can calculate its exact ground state energy~$E_0$ with exact diagonalization.
	A quick recap of the Hubbard model is presented in Section~\ref{sec:hubbard}.
	We performed FCIQMC calculations targeting an increasing number of walkers from $N=100$ to $N=3200$, and plotted both the shift estimator $\overline{S}$ and the uniform projected energy estimator $E^\pm$.
	The results are obtained with time step $\Delta \tau= 10^{-3}$  using $16\times10^6$ iterations with an additional $10^5$ iterations of thermalization.
	The shift is updated every $10$ iterations with damping parameter $\gamma = 0.1$.
	The error bars are calculated using the blocking method of Ref.~\cite{Flyvbjerg89}.
	From this plot, we see that both the projected energy and the shift estimators have a bias that scales down the average number of walkers as $1/\overline{N}$.
	We also note that the shift estimator has a noticeably higher bias than the projected energy at any specific number of walkers.
	The scaling behavior is discussed in the next subsection, while the discrepancy between the different estimators is discussed subsequently.
	
	\subsection{Scaling with the Number of Walkers}\label{sec:scaling}
	To show that the population control bias scales as $1/\overline{N}$, we need to show that the relative bias $B_i$ is independent of the average number of walkers.
	Our starting point is the shift update formula~\eqref{eq:control} relating the shift to the total number of walkers.
	We will assume that the shift is updated continuously to simplify the analysis.
	Nevertheless, the conclusion regarding scaling still holds in the case of a periodic update and is discussed in Appendix~\ref{app:delay}.
	
	By rearranging the terms in the update formula, one can see that the quantity
	\begin{equation}\label{eq:invariant}
	S(\tau) + \frac{\gamma}{A \Delta \tau} \log N(\tau) 
	\end{equation}	
	stays constant for all times $\tau$. Therefore, its covariance with any other statistical quantity vanishes, which allows us to express the covariance between the shift and the population of determinant $\ket{D_i}$ as follows:
	\begin{align}\label{eq:cov}
	\operatorname{Cov}\left(N_i, S\right) &= - \frac{\gamma}{A \Delta \tau} \operatorname{Cov}\left(N_i, \log N\right)\nonumber \\
	&\approx  - \frac{\gamma}{A \Delta \tau} \operatorname{Cov}\left(N_i, \log \overline{N} + \frac{N- \overline{N}}{\overline{N}} \right) \nonumber\\
	&= - \frac{\gamma}{A \Delta \tau } \frac{ \operatorname{Cov}\left(N_i, N\right)}{\overline{N}}\;, 
	\end{align}
	where in the second line, we expanded the logarithm of $N$ around its mean value $\overline{N}$.
	Substituting back in the relative bias definition~\eqref{eq:rel_bias}, we get
	\begin{equation}\label{eq:rel_scaling}
	B_i  \approx  \frac{\gamma}{A \Delta \tau } \frac{ \operatorname{Cov}\left(N_i, N\right)}{\overline{N_i}}   \;.
	\end{equation}
	When there is no sign problem or the number of walkers is above the annihilation plateau, the mean populations of all determinants scale linearly with the total number of walkers.
	Moreover, since the spawning and death/cloning events in FCIQMC are scaled Bernoulli random variables, their variances are proportional to their mean values.
	Therefore, the covariances between the populations of different determinants also scale linearly with the number of walkers.
	As a result, the numerator and denominator of the above ratio have the same scaling, such that the relative bias $B_i$ is mostly independent of the total number of walkers.
	
	\subsection{Different Biases for Different Estimators}\label{sec:discrepancy}
	In Fig.~\ref{fig:scaling}, we saw that the shift estimator has a higher bias than the projected energy estimator at any specific number of walkers.
	We observed this systematic deviation of the two estimators for different systems and was also reported in Ref.~\cite{Vigor2015}.
	To understand it, let us consider Eq.~\eqref{eq:master_biased2} in the ideal case scenario where all determinants have the same relative bias $B_i = B$.
	In that special case, the wavefunction itself and all its associated energy estimators would not be biased, but the average shift would still have the following positive bias:
	\begin{equation}
	\overline{S} - E_0= \frac{B}{\overline{N}}\;.
	\end{equation}
	This suggests that the shift, as an energy estimator, is particularly affected by the population bias.
	
\begin{figure}
		\center
		\includegraphics[width=\columnwidth]{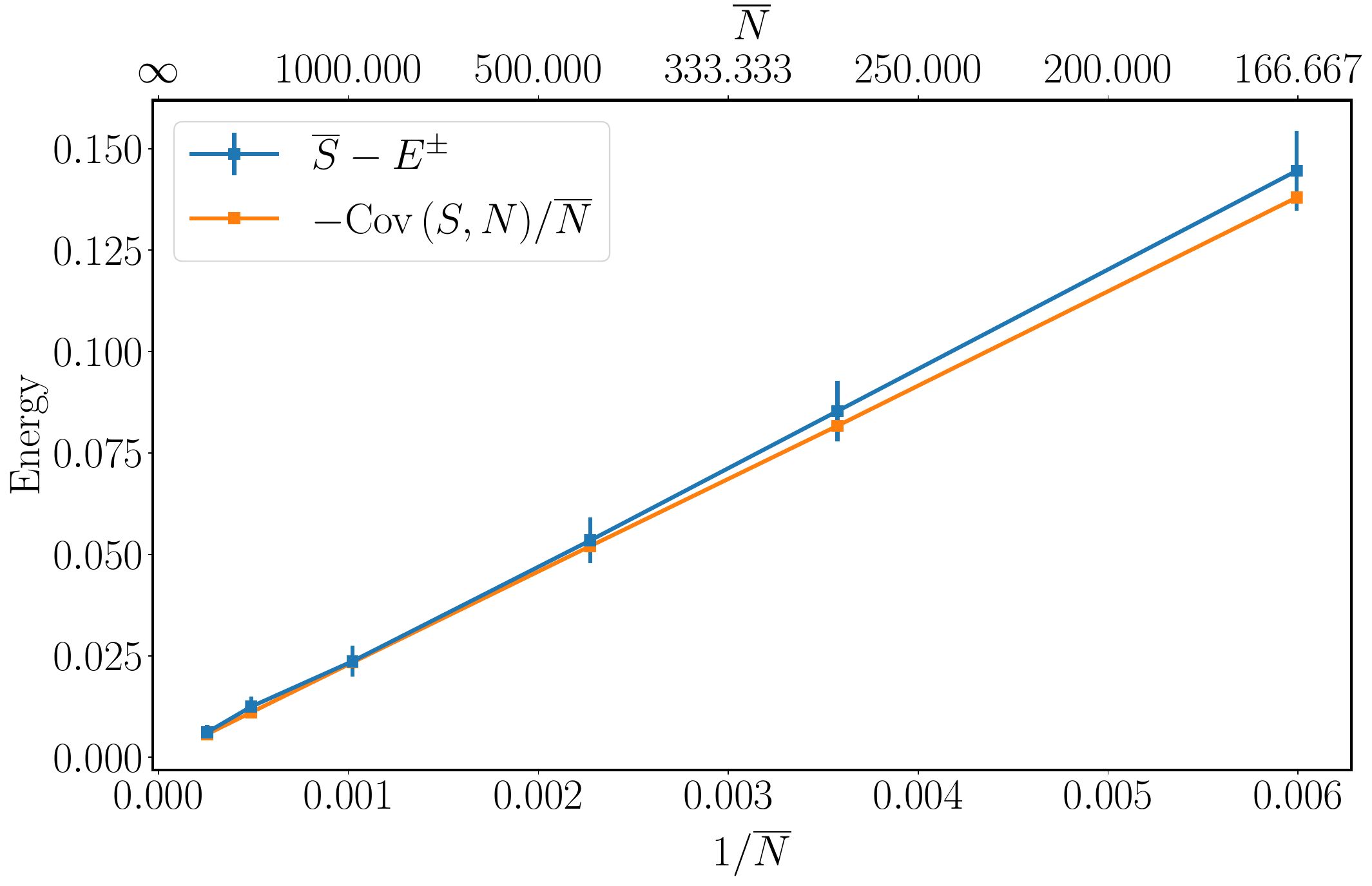}
		\caption{\label{fig:discrepancy}
			A comparison of the two sides of Eq.~\eqref{eq:shift_uinform_pe} for the 14-site one-dimensional Hubbard model of Fig.~\ref{fig:scaling}.
			This relation allows numerically estimating the uniform projected energy $E^\pm$ from the average shift $\overline{S}$ using the average number of walkers $\overline{N}$ and its covariance with shift $\operatorname{Cov}(S, N)$.
		}
	\end{figure}

	We can generally quantify how much the average shift is biased compared to any other projected energy estimator.
	By projecting the biased master equation~\eqref{eq:master_biased} on the corresponding trial wavefunction, we get the time evolution of its overlap with the solution
	\begin{multline}
	-\frac{\d}{\d \tau} \langle\psi^\mathrm{T}\overline{\ket{\psi(\tau)}} = \langle\psi^\mathrm{T}\rvert \bigl(\hat{H}-\overline{S(\tau)}\bigr)\overline{\ket{\psi(\tau)}} \\ - \operatorname{Cov}\bigl[ S(\tau), \braket{\psi^\mathrm{T}|\psi(\tau)}\bigr]
	\end{multline}
	At equilibrium, the left-hand side vanishes and the equation gives a direct relation between the average shift and the projected energy estimate
	\begin{equation}
	\overline{S} - E^\mathrm{T} =  -\frac{\operatorname{Cov}\bigl[S, \braket{\psi^\mathrm{T}|\psi}\bigr]}{\langle\psi^\mathrm{T}\overline{\ket{\psi}}}\;.
	\end{equation}
	This difference in the bias of the two estimators can be easily estimated in FCIQMC using samples of the overlap $\braket{\psi^\mathrm{T}|\psi}$.
	For example, for sign-problem-free cases, the difference between the average shift and the uniform projected energy  estimate can be expressed in terms of the average number of walkers and its covariance with the shift 
	\begin{equation}\label{eq:shift_uinform_pe}
	\overline{S} - E^{\pm} = -\frac{\operatorname{Cov}\left(S, N\right)}{\overline{N}}\;.
	\end{equation}
	In Fig.~\ref{fig:discrepancy}, we show that this equation is indeed satisfied by the earlier results of the Hubbard model.
	This relation provides a simple way of partially correcting the bias of the average shift in cases where calculations of projected energies are not practically feasible.
	To remove the bias completely, we need to correct the wavefunction itself, which we address in the next section.
	It is worth noting that Fig.~\ref{fig:discrepancy} demonstrates that the covariance $\operatorname{Cov}(S, N)$ is mostly independent of $\overline{N}$, which is consistent with the same behavior of $\operatorname{Cov}(S, N_i)$ concluded from the previous section.
	The dependence of the covariance terms on other simulation parameters is discussed in Appendix~\ref{app:cov}.
	
	\section{Bias Correction}
	The intuitive idea behind eliminating population control bias is noting that the role of updating the shift is merely scaling the wavefunction.
	The problem is that the scale factors are correlated with the sampled wavefunctions.
	Nevertheless, if we keep track of these factors and use them to scale the wavefunctions back, we can undo the effects of the shift updates.
	As a result, these rescaled wavefunctions would then satisfy the  Schr\"odinger equation with a constant shift and thus without bias.
	
	In the following subsection, we show mathematically how this is achieved by a reweighting of FCIQMC samples. 
	The reweighting technique can be used to correct the estimates of any wavefunction-related quantity, but it cannot be applied directly to the shift average.
	So we show next how to fix the shift estimator properly.
	Finally, we discuss some technical details of the reweighting technique.
	
	\subsection{Correcting the Wavefunction}
	In FCIQMC, the time derivative in Schr\"odinger equation is discretized using Euler's explicit method, leading to the linear propagator
	\begin{equation}\label{eq:linear_iteration}
	\overline{\ket{\psi(\tau+\Delta \tau)}} = \left[\hat{I}-\Delta \tau \left(\hat{H} - S(\tau)\right) \right]\ket{\psi(\tau)}\;.
	\end{equation}
	However, to make the derivation and results more streamlined, we assume that FCIQMC is using the following exponential propagator instead of its linear approximation
	\begin{equation}\label{eq:exact_iteration}
	\overline{\ket{\psi(\tau+\Delta \tau)}} = \exp\left[-\Delta \tau \left(\hat{H} - S(\tau)\right) \right]\ket{\psi(\tau)}
	\end{equation}
	This replacement of the linear propagator by the exponential one does not affect the validity of the results.
	The reason is that the next step requires decomposing the propagator into the product of two propagators.
	For the linear propagator, this decomposition introduces a time-step error of the order $\mathcal{O}(\Delta \tau^2)$, which is the same order of error introduced by using the exponential propagator instead.
	
	Let us now decompose the propagator into one with a hypothetical constant shift $C$ and a remaining pure scaling term
	\begin{multline}
	\exp\left[-\Delta \tau \left(\hat{H} - S(\tau)\right) \right] 	 = \exp\left[-\Delta \tau \left(\hat{H} - C\right) \right] \times\\  \exp\left[-\Delta \tau \left( C \vphantom{\hat{H}}- S(\tau) \right) \right] 
	\end{multline}
	Substituting back in Eq.~\eqref{eq:exact_iteration}, we get the following equation, which is (approximately) satisfied at every iteration of FCIQMC
	\begin{multline}
	\exp\left[\Delta \tau \left( C\vphantom{\hat{H}} - S(\tau)\right) \right]  \overline{\ket{\psi(\tau+\Delta \tau)}} =  \\ \exp\left[-\Delta \tau \left(\hat{H} - C\right) \right]   \ket{\psi(\tau)}
	\end{multline}
	Denoting $X_C(\tau)  \coloneqq \exp\left[\Delta \tau \left( C\vphantom{\hat{H}} - S(\tau)\right) \right] $ and taking the average of both sides
	\begin{equation}
	\overline{X_C(\tau)  \ket{\psi(\tau+\Delta \tau)}} =   \exp\left[-\Delta \tau \left(\hat{H} - C\right) \right]  \overline{\ket{\psi(\tau)}}\;,
	\end{equation}
	we see that the average of $X_C(\tau)  \ket{\psi(\tau+\Delta \tau)}$ is the \emph{unbiased} evolution of the average of $\ket{\psi(\tau)}$ using  the constant shift $C$ .
	
\begin{figure}
		\center
		\includegraphics[width=\columnwidth]{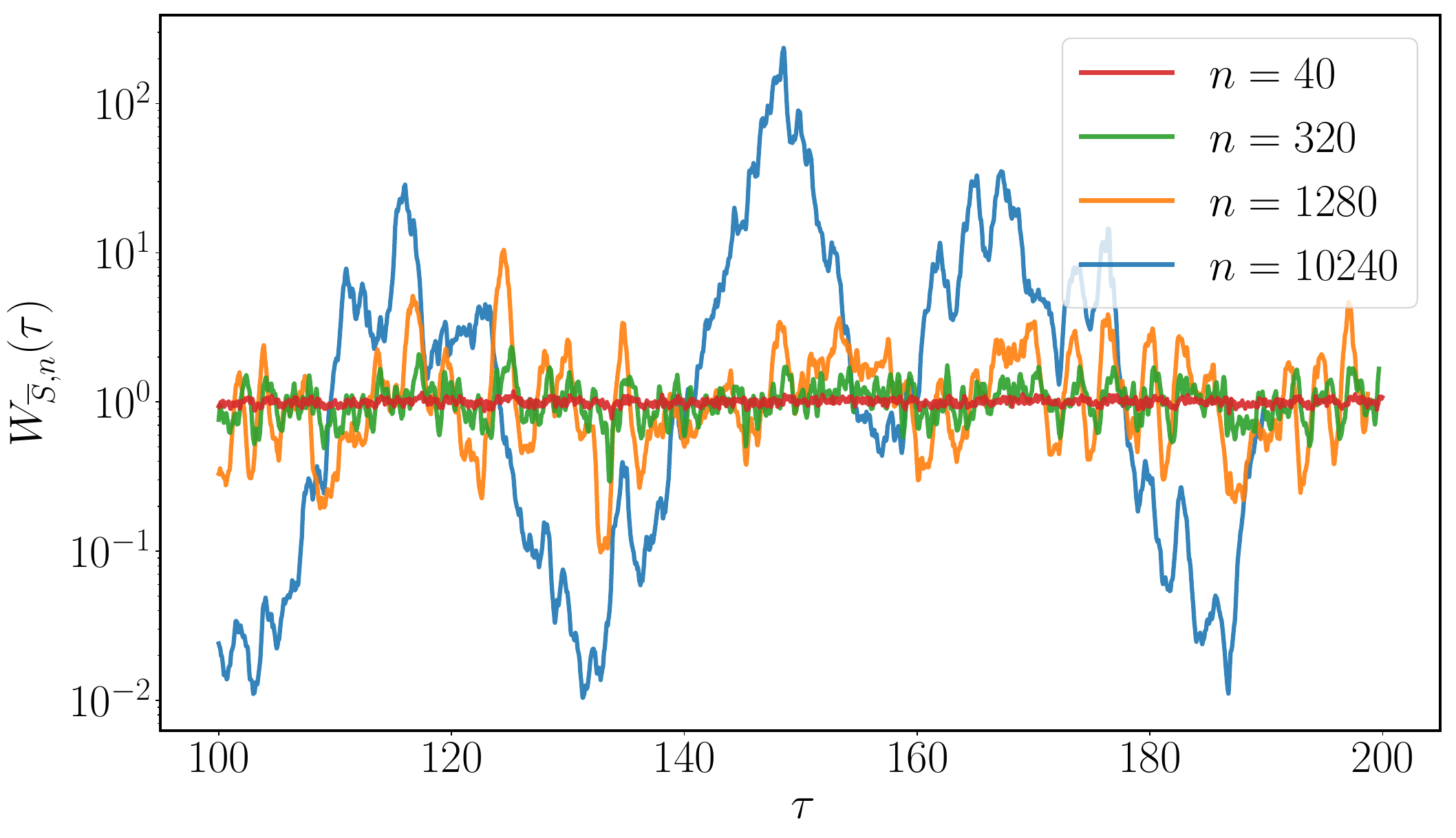}
		\caption{\label{fig:weights}
			Correction weights for the 14-site Hubbard model using different correction orders $n$.
			The data corresponds to an average number of walkers $\overline{N}=167$.
		}
	\end{figure}

	By repeating the above analysis for an arbitrary number of iterations $n$, we get the following general relation
	\begin{equation}\label{eq:correction}
	\overline{W_{C,n}(\tau)  \ket{\psi(\tau)}} =    \exp \left[-\Delta \tau \left(\hat{H} - C\right) \right]^n   \overline{\ket{\psi(\tau-n\Delta \tau)}}\;,
	\end{equation}
	where the weight factor is now the product of subsequent scaling factors
	\begin{align}\label{eq:weights}
	W_{C,n}(\tau) &\coloneqq \prod_{j=1}^{n} X_C(\tau-j\Delta \tau) \nonumber \\
	&= \exp\left[-\Delta \tau \sum_{j=1}^{n} \left(S(\tau-j\Delta \tau) -C\vphantom{\hat{H}}\right)\right]\;.
	\end{align}
	This expression shows that by performing enough iterations and using a large-enough value of $n$, we can project out the exact ground state $\ket{\phi_0}$ without any population control bias by reweighing the wavefunction at iteration $\tau$ with weight  $W_{C,n}(\tau)$
	\begin{equation}\label{eq:limit}
	\lim_{n\to\infty} \overline{W_{C,n}(\tau)  \ket{\psi(\tau)}} \propto \ket{\phi_0}\;.
	\end{equation}
	In Fig.~\ref{fig:weights}, we plot these weights for the 14-site Hubbard model using different values of $n$.
	The value $n$ can be thought of as the \emph{order of correction} applied to the wavefunction.
	Weights of order $n$ reflect how the amplitude of the wavefunction would change if the population control had been suspended for a period of $n \Delta \tau$.
	For large values of $n$, the weights fluctuate over several orders of magnitude, which effectively reduces the number of samples in any averaged quantity.
	This puts a practical limit on the chosen value of $n$, as we will see later.
	
	\subsection{Correcting Energy Estimators}
	Correcting the projected energy estimators using the wavefunction weights is straightforward.
	One has to reweight the terms appropriately in both the numerator and the denominator of Eq.~\eqref{eq:pe_estimate}
	\begin{equation}
	E^\mathrm{T}_\text{corrected} \approx \frac{\sum_{i=1}^L W_{C, n}(\tau_i)  \braket{\psi^\mathrm{T}|H|\psi(\tau_i)}}{\sum_{i=1}^L W_{C, n}(\tau_i) \braket{\psi^\mathrm{T}|\psi(\tau_i)}}\;.
	\end{equation}
	To correct the shift, however, the reweighting cannot be applied directly.
	We need first to relate the shift to the corrected wavefunction.
	Being able to correct the shift is valuable because, for some systems, obtaining reliable projected energy estimates is computationally expensive while the shift is readily available.	
	
	Evolving the exact ground state wavefunction $\ket{\phi_0}$ using  the Schr\"odinger equation with fixed shift $C$ leads to an exponential decay of its norm with rate $E_0-C$
	\begin{equation}
	\exp\left[-\Delta \tau \left(\hat{H} - C\right) \right]  \ket{\phi_0} = \exp\left[-\Delta \tau \left(E_0 - C\right)\right] \ket{\phi_0} 
	\end{equation}
	If we project the above equation on some trial state $\ket{\psi^\mathrm{T}}$ and rearrange the terms we get what is known as the \emph{growth estimator} of the ground state energy
	\begin{equation}\label{eq:growth}
	E^\mathrm{T}_\text{growth} = C - \frac{1}{\Delta \tau } \log\left[\frac{\braket{\psi^\mathrm{T}|\exp\left[-\Delta \tau \left(\hat{H} - C\right) \right] |\phi_0}}{\braket{\psi^\mathrm{T}|\phi_0}}\right]
	\end{equation}
	We can approximate both the numerator and denominator using equation \eqref{eq:correction} and a large enough value of~$n$
	\begin{align}
	\ket{\phi_0} &\approx  \overline{W_{C,n}(\tau)  \ket{\psi(\tau)}}  \\
	\exp\left[-\Delta \tau \left(\hat{H} - C\right) \right]  \ket{\phi_0}  &\approx  \overline{W_{C,n+1}(\tau+\Delta \tau)  \ket{\psi(\tau+\Delta \tau)}} 
	\end{align}
	Substituting back in Eq.~\eqref{eq:growth}, and replacing the mean values by averages over $L$ iterations, we get
	\begin{equation}
	E^\mathrm{T}_\text{growth} \approx C - \frac{1}{\Delta \tau } \log\left[\frac
	{\sum_{i=1}^{L} W_{C,n+1}(\tau_{i+1}) \braket{\psi^\mathrm{T} | \psi(\tau_{i+1})}}
	{\sum_{i=1}^{L} W_{C,n}(\tau_i) \braket{\psi^\mathrm{T} | \psi(\tau_i)}}
	\right]
	\end{equation}
	This expression is valid for any value of $C$ below the first excitation energy and any trial wavefunction $\psi^\mathrm{T}$ with nonzero overlap with the ground state.
	The specific choices may only influence the statistical uncertainty of the estimator.
	
\begin{figure}
		\center
		\includegraphics[width=\columnwidth]{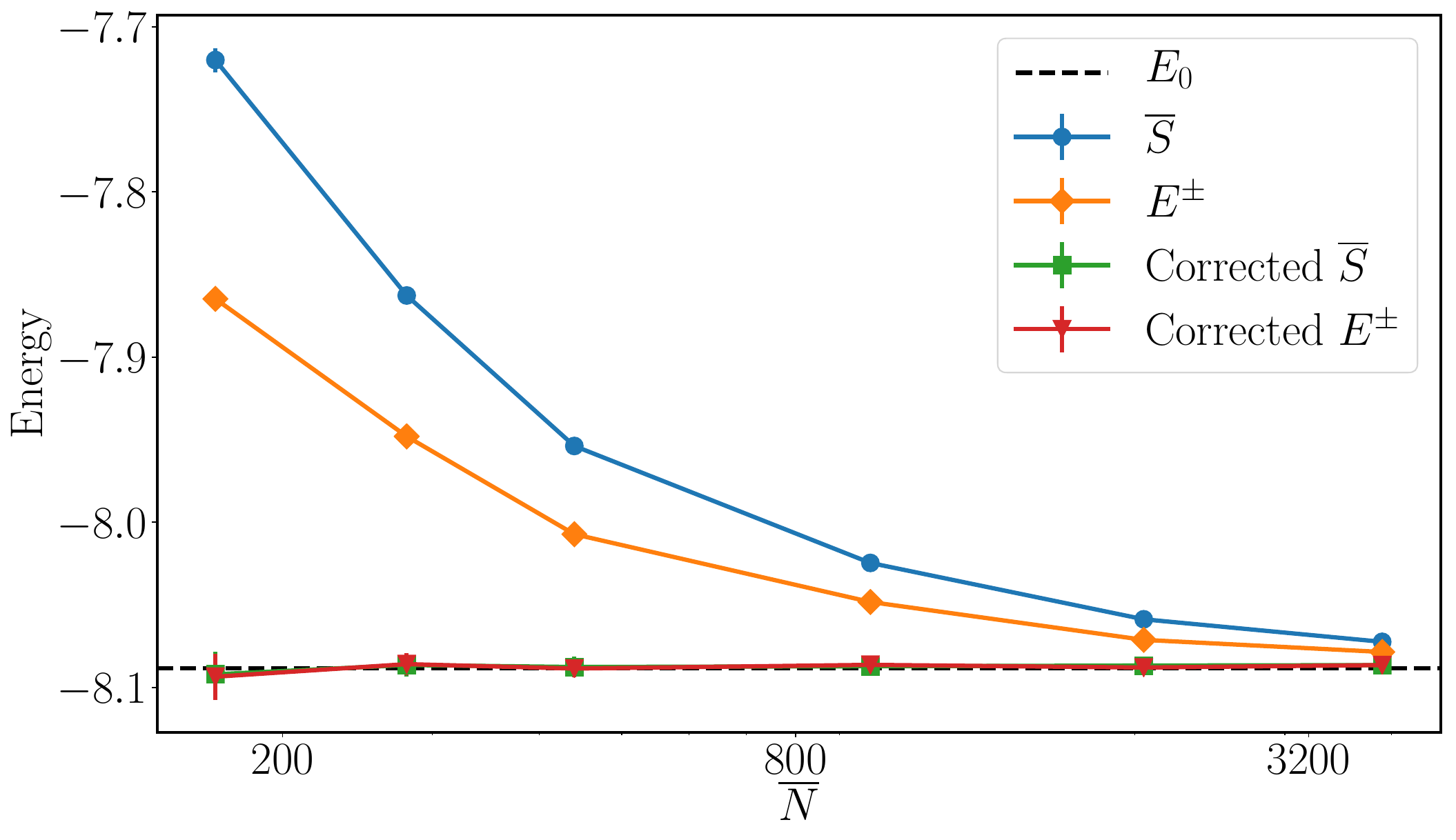}
		\caption{\label{fig:correction}
			Corrected energy estimates of the 14-site Hubbard model.
			All points have error bars, but most of them are too small to be seen on this plot.
			Within these error bars, the corrected shift $\overline{S}$ and the corrected uniform projected energy $E^\pm$ agree with each other and with the exact energy $E_0$.
		}
	\end{figure}

	Incidentally, by setting $C$ to the average shift $\overline{S}$, the second term gives us the necessary correction to remove the population control bias of the average shift.
	In fact, using the average shift as $C$ is optimal in the sense that its value minimizes the squared distance to all sampled shift values and thus minimizes the fluctuations of the scale factors $X_C$.
	As was the case for project energies, the choice of the trial wavefunction depends on the system.
	It is reasonable in molecular systems to use the reference determinant or the ground state of a trial space constructed out of the most populated determinants.
	In sign-problem-free systems, using the uniform trial wavefunction of Eq.~\eqref{eq:uniform_trial} the overlap $\braket{\psi^\mathrm{T} | \psi(\tau)}$ is just the total number of walkers $N(\tau)$
	\begin{equation}\label{eq:corrected_shift}
	\overline{S}_\text{corrected} \approx \overline{S} - \frac{1}{\Delta \tau } \log\left[\frac
	{\sum_{i=1}^{L} W_{\overline{S},n+1}(\tau_{i+1}) N(\tau_{i+1})}
	{\sum_{i=1}^{L} W_{\overline{S},n}(\tau_i) N(\tau_{i})}
	\right]
	\end{equation}
	In Fig.~\ref{fig:correction}, we plot the results of applying this reweighting technique to the 14-sties Hubbard model.
	The reweighting eliminates the population control bias in both the shift and projected energy estimates, with the corrected values falling spot on the exact energy.
	In this plot, we used $n=2560$ terms in the weight factors.
	
	\subsection{Correction Order}
\begin{figure}
		\center
		\includegraphics[width=\columnwidth]{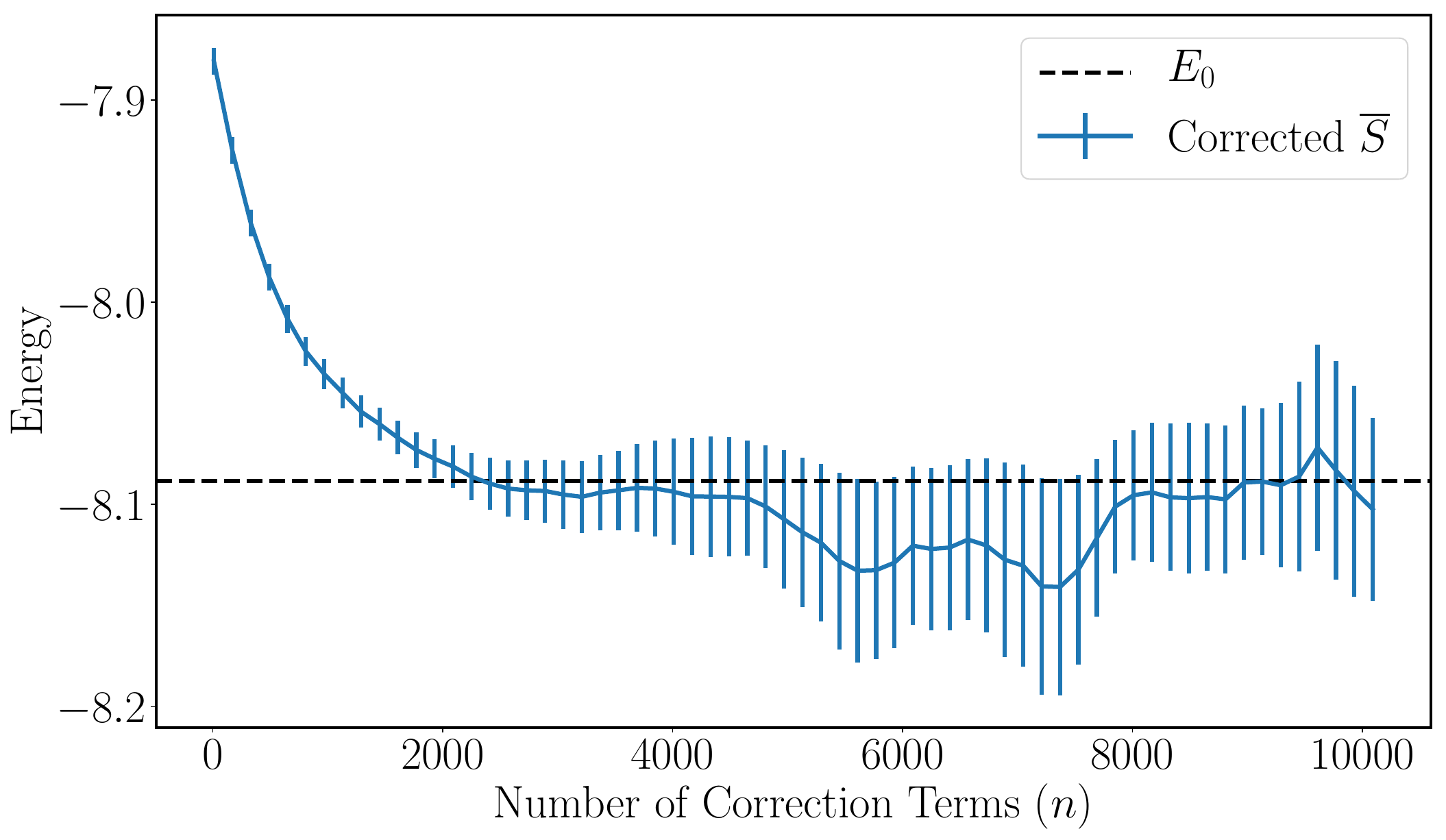}
		\caption{\label{fig:order}		
			Corrected shift estimates with increasing orders of correction $n$ for the 14-site  Hubbard model.
			The data corresponds to the average number of walkers $\overline{N}=167$.
		}
	\end{figure}

\begin{figure}
		\center
		\includegraphics[width=\columnwidth]{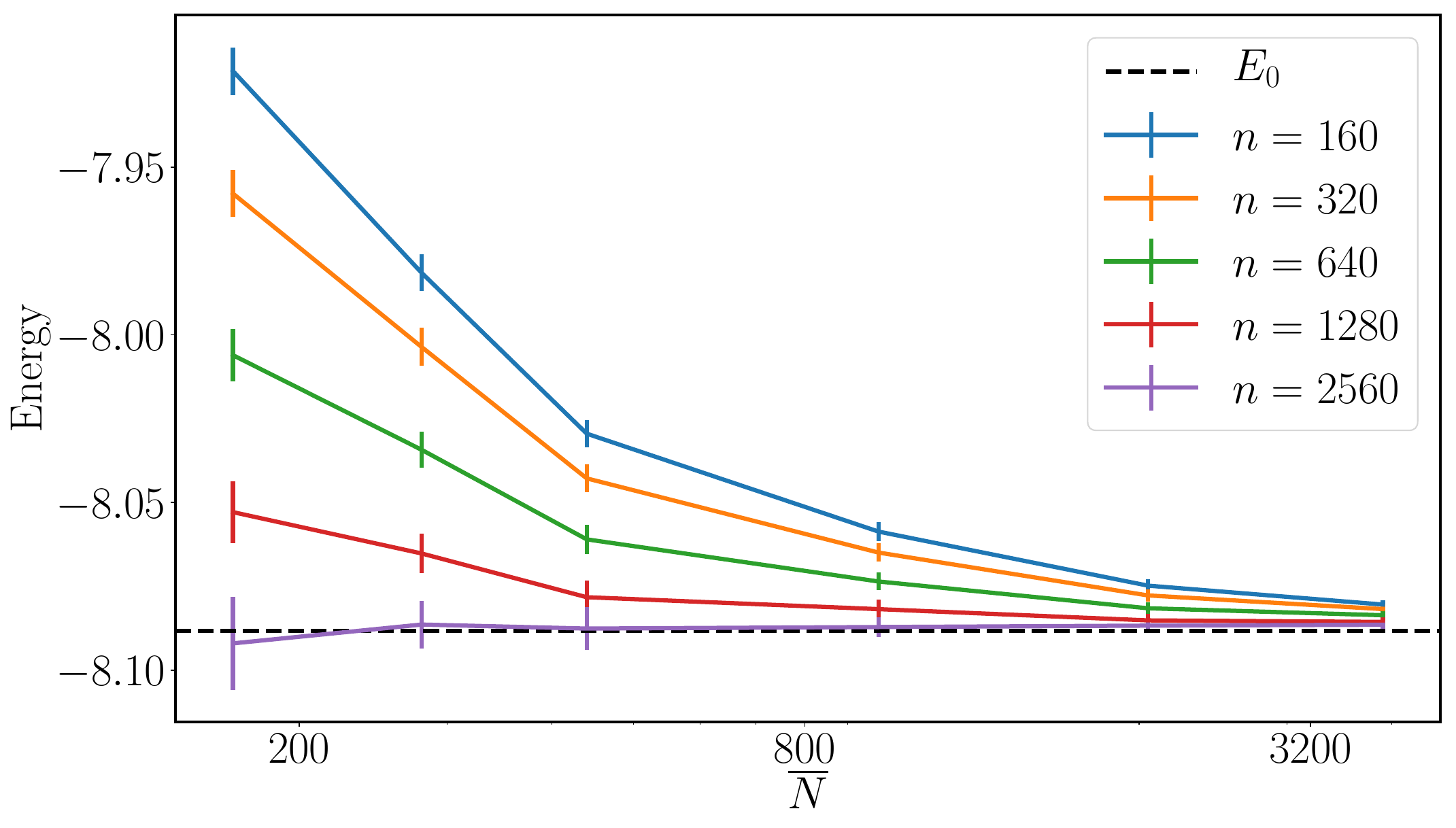}
		\caption{\label{fig:walkers_order}
			Corrected shift estimates of the 14-site Hubbard model using different orders of correction  $n$ and different average number of walkers $\overline{N}$.
		}
	\end{figure}

	The number of correction terms $n$ included in the weight factors $W_{C, n}$ affects both the accuracy and precision of the corrected energy estimates.
	Ideally, we want to include as many terms as possible because more terms imply longer projection times, Eq.~\eqref{eq:correction}, thus further reduction of the population control bias (better precision).
	On the other hand, including more terms implies longer times without population control and larger the fluctuations of the weight factors, Fig.~\ref{fig:weights}, thus larger error bars (less accuracy).
	In Fig.~\ref{fig:order}, we plot the corrected shift estimates against the correction order $n$.
	As the correction order increases, the population control bias decays exponentially while the statistical error bars get larger.
	After some point, around $n =3000$, increasing the correction order has diminishing returns for the bias but still increases the error bars.
	The optimal value of $n$ is the one for which the bias is just below the statistical uncertainty.
	Knowing this value would not be possible without knowing the exact result.
	As a heuristic, we suggest choosing $n$ as the lowest value, after which the energy estimate goes up (assuming the bias is positive).
	When a set of calculations with different numbers of walkers is available, a better choice can be made.
	Choose the value for which the corrected energy estimates from different number walkers agree within the error bars.
	For example, this is the value $n=2560$ for the result of the 14-site Hubbard shown in Fig.~\ref{fig:walkers_order}.
	
	Finally, it is worth pointing out that when comparing the corrected estimates of different calculations using different time steps $\Delta \tau$, the relevant quantity is the total imaginary time  $T \coloneqq n \Delta \tau$, which should be long enough to project out the bias and make it smaller than the statistical error bars.
	Consequently, it is desirable to choose the time step as large as possible.
	On the other hand, the correction formula has a local truncation error of the order $\mathcal{O}(\Delta \tau^2)$ due to the replacement of the linear propagator by the exponential one. 
	The global truncation error thus scales as $\mathcal{O}(\Delta \tau)$ for a fixed projection time $T$.
	In practice, however, we did not observe this error in our calculations even when the time-step was close to the maximum allowed value ${2 /(E_\text{max}-E_0)}$, where $E_\text{max}$ is the maximum eigenvalue of the Hamiltonian.
	
	\section{Role of Importance Sampling} \label{sec:imp_samp}
	Since the population control bias is proportional to the covariance between the shift and the wavefunction, reducing the variance of the shift or the wavefunction should help reduce the bias.
	A well-known technique for reducing the variance of quantum Monte Carlo estimates is importance sampling~\cite{Kalos1974, Umrigar2015}.
	Given a guiding wavefunction $\ket{\psi_{\mathrm{G}}}$, importance sampling  corresponds to solving  an auxiliary problem with the modified Hamiltonian
	\begin{equation}
	\tilde{H}_{ij} = H_{ij} \frac{\braket{D_i| \psi_{\mathrm{G}}}}{\braket{D_j| \psi_{\mathrm{G}}}}\;.
	\end{equation}
	The wavefunctions of the original and auxiliary problem are then related by
	\begin{equation}
	\tilde{N_i} = N_i \braket{D_i| \psi_{\mathrm{G}}}\;.
	\end{equation}
	Using a guiding wavefunction that is close to the ground state, reduces the fluctuation in the number of walkers and the shift.
	The reason is that a walker on a determinate $\ket{D_j}$ contributes \emph{on average}  $1+\Delta \tau (\sum_i \tilde{H}_{ij} -S)$ walkers to the next iteration.
	As the guiding wavefunction becomes closer to the ground state, the variation in the column-sum $\sum_i \tilde{H}_{ij} $ gets smaller, reducing the variation in the number of walkers and its dependence on the details of the sampled wavefunction.
	From a different perspective, the propagator becomes closer to a column-stochastic matrix alleviating the need for population control. Importance sampling has also been discussed in the context of Density Matrix Quantum Monte Carlo \cite{Blunt2014} and used in projective quantum Monte Carlo simulations of Ising chains~\cite{Inack2018}.
	
	As an illustrative example, take the limiting case where the guiding wavefunction is the exact ground state.
	Then the column-sum of the modified Hamiltonian matrix elements becomes a constant
	\begin{equation}
	\sum_i \tilde{H}_{ij} = \sum_i  H_{ij} \frac{\braket{D_i| \phi_0}}{\braket{D_j| \phi_0}}  =  E_0 
	\end{equation}
	and thus all walkers have the same reproduction rate ${1+\Delta \tau (E_0 -S)}$  .
	Moreover, the uniform projected energy $E^\pm$ equals the grounds state energy $E_0$ regardless of the sampled wavefunction.
	Therefore, the population control bias is given entirely by Eq.~\eqref{eq:shift_uinform_pe} which only depends on the covariance of the shift with the number of walkers rather than with the details of the sampled wavefunction.

	\begin{figure}
		\center
		\includegraphics[width=\columnwidth]{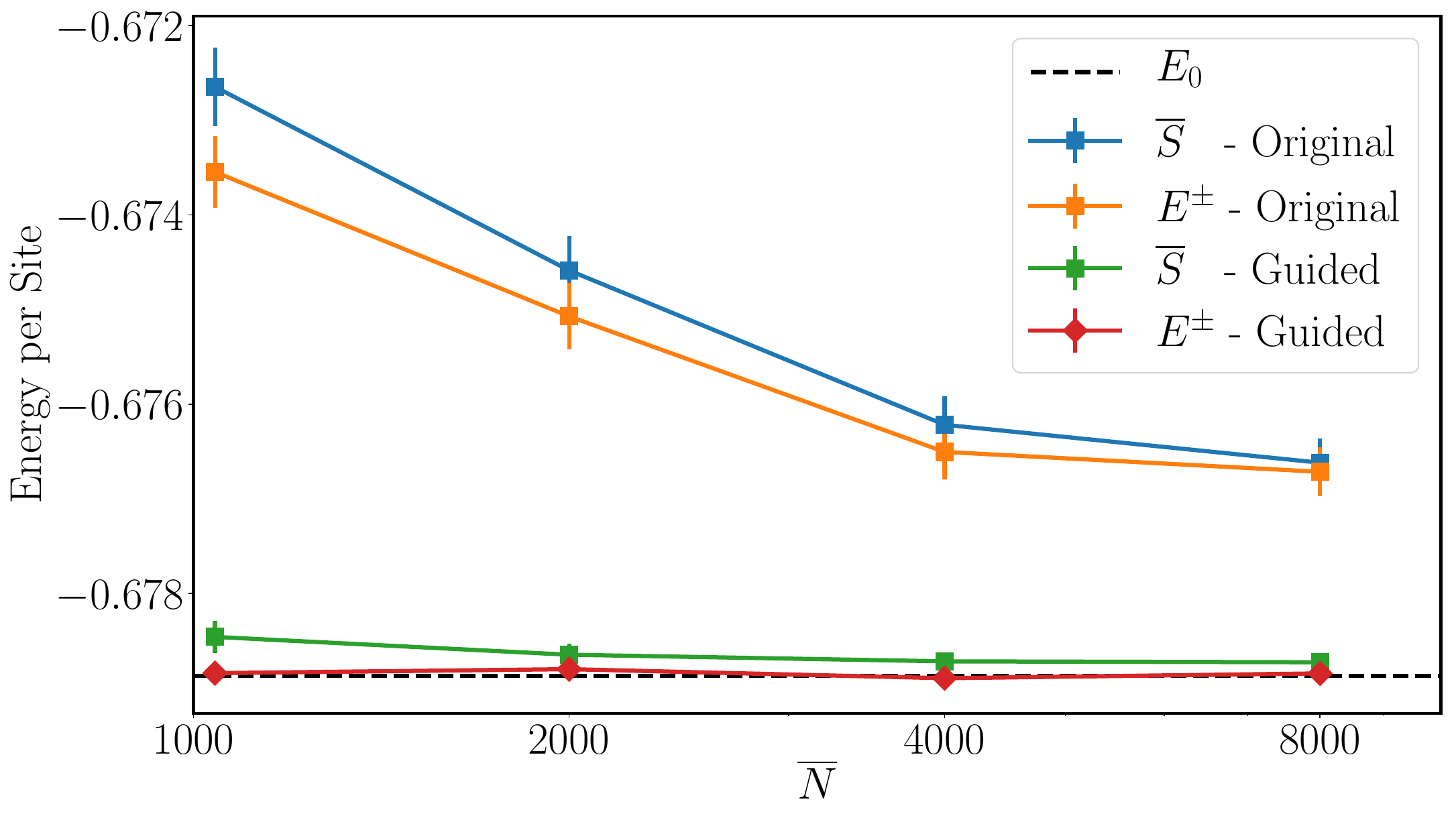}
		\caption{\label{fig:importance}
			Comparison of the energy estimates of the $6\times 6$ Heisenberg model with and without importance sampling (labeled ``Guided'' and ``Original'', respectively).
			The importance sampling uses a Gutzwiller wavefunction with parameter $g=3$.
		}
	\end{figure}
	
	\begin{figure}
		\center
		\includegraphics[width=\columnwidth]{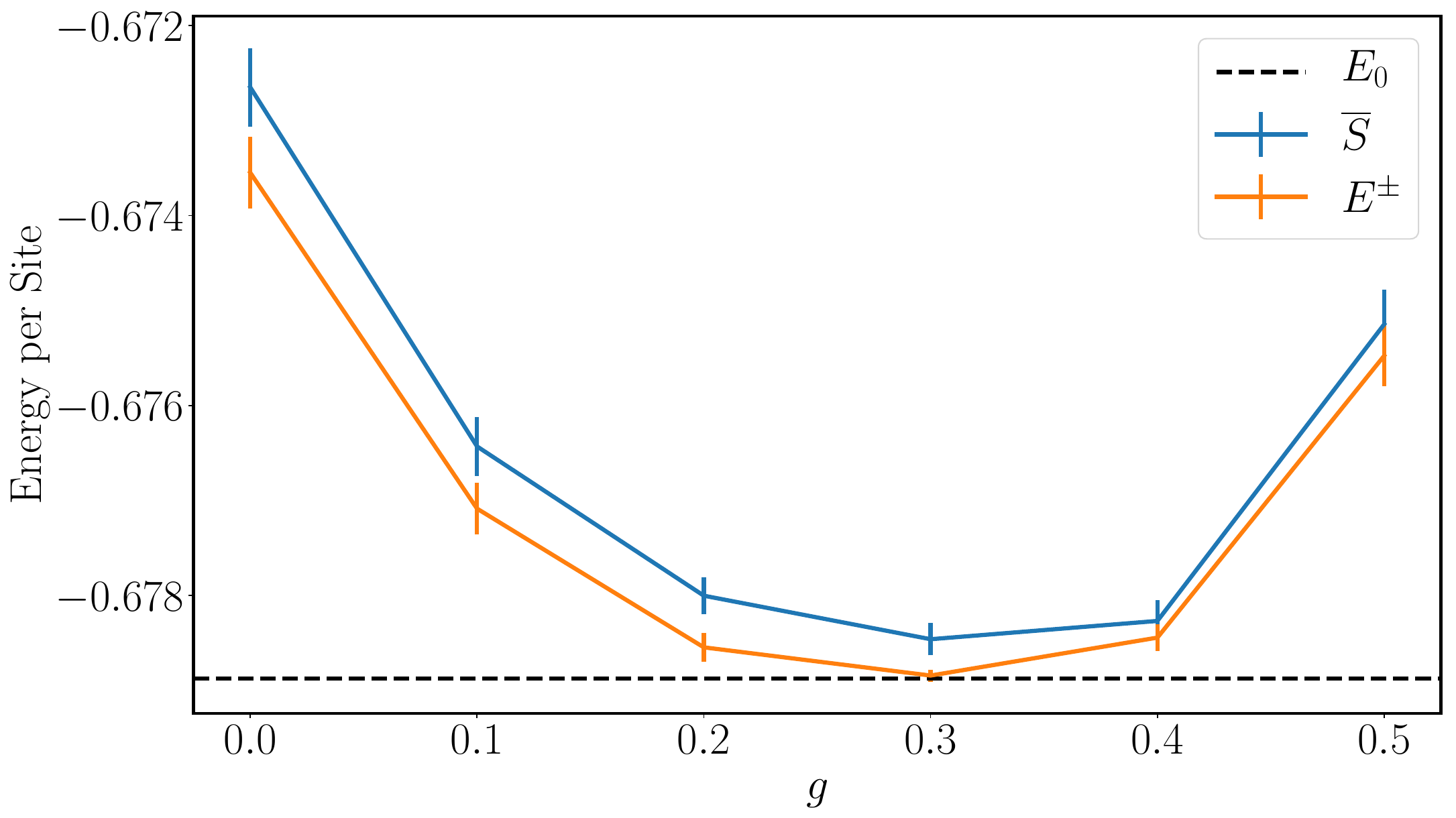}
		\caption{\label{fig:optimize}
			FCIQMC energy estimates of the $6\times 6$ Heisenberg model using and $1000$ walkers and different values of the Gutzwiller wavefunction parameter $g$.
		}
	\end{figure}
	In practice, the guiding wavefunction does not necessarily need to be very close to the ground state.
	Using even the simplest forms of guiding wavefunctions can have a noticeable impact on the population control bias.
	In this paper, we employ the following Gutzwiller-like wavefunction~\cite{Gutzwiller1963}, where energetically-unfavorable determinants are exponentially suppressed
	\begin{equation}\label{eq:gutzwiller}
	\ket{\psi_\text{G}} = \sum_i \e^{-g H_{ii}} \ket{D_i}\;,
	\end{equation}
	and $g$ is a non-negative parameter to be optimized later.
	Implementing importance sampling using this guiding wavefunction is straightforward in FCIQMC.
	Spawns from determinant $\ket{D_i}$ to determinant $\ket{D_j}$ have to be scaled by the factor
	\begin{equation}
	w = \exp\bigl[-g(H_{jj}-H_{ii})\bigr]\;.
	\label{eq:simple_gutzw}
	\end{equation}
	Computing this factor incurs only a small computational overhead, and for local lattice models, it can be evaluated in a way that is independent of the system size.
	In Fig.~\ref{fig:importance}, we show the effect of using this guiding wavefunction on the $6\times 6$ Heisenberg model with periodic boundary conditions (see Section~\ref{sec:heisenberg}).
	The plot demonstrates how markedly-effective the importance sampling is in reducing not only the statistical error bars but also the population control bias of both the shift and the projected energy estimators.

	The guiding wavefunction parameters can be optimized by minimizing the variational energy as in variational Monte Carlo or by solving an eigenvalue equation in a projected subspace as done in coupled-cluster methods.
	Besides these two options, we found that a straightforward bootstrapping method works very well for the sake of minimizing population control bias. 
	We first optimize the parameters using low-accuracy but cheap FCIQMC energy estimates and later use these optimal parameters to obtain a highly-accurate energy estimate employing a much larger number of walkers.
	For example, in the earlier case of the Heisenberg model, the parameter $g$ of Gutzwiller wavefunction was chosen by running different FCIQMC calculation using $1000$ walkers and different values of $g$ and then selecting the one giving the lowest energy estimates as shown in Fig.~\ref{fig:optimize}.

	\section{Applications}\label{sec:applications}
	{ 
		\renewcommand{\arraystretch}{1.5}
		\begin{table*}[t]
			\begin{tabular}{@{\extracolsep{6pt}}lllll}
				System & Sites & Method & Energy per site (PBC) & Energy per site (OBC) \\ \hline \hline
				\multirow{5}{*}{\shortstack[l]{$U/t = 4$\\ half-filling}} & \multirow{4}{*} {102} 
				& DMRG & $-0.573~79$ & $-0.570~13$\\
				\cline{3-5}
				& & FCIQMC -- Original & $-0.570~76(40)$ & $-0.566~80(50)$\\
				& & FCIQMC -- Guided & $-0.573~71(8)$ & $-0.570~11(7)$\\
				& & FCIQMC -- Corrected & $-0.573~75(8)$ & $-0.570~17(9)$\\
				\cline{2-5}
				& $\infty$ & Bethe ansatz & \multicolumn{2}{c}{$-0.573~73$}\\
				\hline\hline
				\multirow{9}{*}{\shortstack[l]{$U/t = 8$\\ half-filling}} & \multirow{4}{*} {102}
				& DMRG & $-0.327~57$ & $-0.325~50$\\
				\cline{3-5}
				& & FCIQMC -- Original & $-0.322~99(49)$ & $-0.321~19(43)$\\
				& & FCIQMC -- Guided & $-0.327~54(1)$ & $-0.325~48(2)$\\
				& & FCIQMC -- Corrected & $-0.327~55(3)$ & $-0.325~49(3)$\\
				\cline{2-5}
				& \multirow{4}{*} {150}
				& DMRG & $-0.327~55$ & \\
				\cline{3-5}
				& & FCIQMC -- Original & $-0.302~54(154)$ & \\
				& & FCIQMC -- Guided & $-0.327~39(6)$ & \\
				& & FCIQMC -- Corrected & $-0.327~54(9)$ &\\
				\cline{2-5}
				& $\infty$ & Bethe ansatz & \multicolumn{2}{c}{$-0.327~53$}\\
				\hline\hline
				\multirow{4}{*}{\shortstack[l]{$U/t = 8$\\ 4 holes ($M_s = 0$)}} & \multirow{4}{*}{102}
				& DMRG & $-0.392~29$& $-0.390~04$\\
				\cline{3-5}
				& & FCIQMC -- Original & $-0.388~52(32)$ & $-0.387~09(39)$\\
				& & FCIQMC -- Guided & $-0.392~28(3)$ & $-0.390~02(2)$\\
				& & FCIQMC -- Corrected & $-0.392~29(3)$ & $-0.390~03(2)$\\
				\hline\hline
			\end{tabular}
			\caption{\label{table:hubbard}
				Results for the 1D Hubbard model. Technical details for the FCIQMC calculations are shown in Table~\ref{table:hubbard_fciqmc_details} and for the DMRG calculations in Table~\ref{table:hubbard_dmrg_bond_dims}, respectively. ``FCIQMC -- Original'' implies no bias correction or importance sampling. ``FCIQMC -- Guided'' implies the use of importance sampling with the Gutzwiller-like factor. ``FCIQMC -- Corrected'' implies the population control bias correction on top of the importance sampling. Chain lengths of $\infty$ indicate calculations in the thermodynamic limit.
			}
			
		\end{table*}
	}

	To demonstrate the capabilities of the presented improvements of the FCIQMC algorithm, we apply it to sign-problem-free lattice models with large lattice sizes.
	The calculations were performed using the efficient NECI implementation~\cite{NECI}.
	In these calculations, we employed importance sampling using the Gutzwiller-like wavefunction of Eq.~\eqref{eq:gutzwiller} and optimized its parameter $g$ as described earlier.
	Since calculating the uniform projected energy gets more expensive for larger grid sizes, we relied solely on the shift as an energy estimator.
	To account for any remaining population control bias, we used the reweighting formula of Eq.~\eqref{eq:corrected_shift} to correct the shift average.
	
	\subsection{Hubbard Model}\label{sec:hubbard}
	The Hubbard Hamiltonian on a general lattice reads
	\begin{equation}
	\hat{H} = -t \sum_{\langle ij\rangle, \sigma} \hat{c}^\dagger_{i, \sigma} \hat{c}_{j, \sigma} + U \sum_{i} \hat{n}_{i,\uparrow}  \hat{n}_{i,\downarrow}
	\end{equation}
	where $\langle ij \rangle$ indicates summation over all neighboring sites, $\hat{c}^\dagger_{i, \sigma} (\hat{c}_{i, \sigma}) $ is the fermionic creation (annihilation) operator of an electron of spin $\sigma$ at site $i$ and $ \hat{n}_{i,\sigma} = \hat{c}^\dagger_{i, \sigma} \hat{c}_{i, \sigma}$ is the corresponding number operator, $t$ is the hopping amplitude between nearest neighbors and $U$ is on-site Coulomb repulsion.
	This simple model of interacting electrons on a lattice has been under study for over 50 years and is used as a model for understanding the physics of high-temperature superconductors~\cite{hubbard_electron_1963, Gutzwiller1963, kanamori_electron_1963, Zhang88}.
	Interestingly, the one-dimensional case can be solved in the thermodynamic limit by employing Bethe ansatz, which reduces the problem to a set of coupled algebraic equations, Lieb--Wu equations, which can be solved analytically in the thermodynamic limit~\cite{Lieb68, Takahashi72}.
	
	Using open boundary conditions, the 1D Hubbard model does not have a sign problem in FCIQMC.
	This can be seen by ordering the orbitals by their spin then by their lattice site. Since in real space, it is only kinetic energy terms that cause transitions between two Slater determinants (by hopping single electrons between nearest-neighbor sites), and the sign of these matrix elements are negative, and in addition, since (in 1D and open boundary conditions) there is no possibility for an electron to return to its original lattice site via alternative routes, the problem is sign-problem-free.
	With periodic boundary conditions, electrons hopping across the boundary can, in general, lead to positive matrix elements due to a possible negative Fermi-sign from the paths that wrap around the boundary and return to the original Slater determinant.
	Nevertheless, if the number of spin-up electrons and the number of spin-down electrons are both odd, the number of required commutations will always be even, and thus no Fermi-sign arises. This means that specific off-half-filling periodic models can also be studied with the FCIQMC method without any sign problem.
	
	In the following, we look at some paradigmatic long 1D Hubbard systems.
	We benchmark our results against DMRG results obtained using the BLOCK code~\cite{chan_introduction_2008}. DMRG is a deterministic method that uses the so-called \emph{matrix product state (MPS)} to represent the ground-state wavefunction. The MPS is iteratively optimized in sweeps through the sites in a predefined order, which is trivially chosen in 1D systems. The method benefits from the low-entanglement of the ground states that is especially present in 1D problems. The computational resources required to solve a system with DMRG up to a given accuracy is mainly determined by the \emph{bond dimension} $M$. DMRG also benefits from the locality, which is broken by introducing periodic boundary conditions. 
	
	In Table~\ref{table:hubbard}, we show FCIQMC and DMRG results of the 102-site Hubbard chain for $U/t = 4$ and~$8$ at half-filling as well as the 150-site Hubbard chain at $U/t = 8$. 
	These results are also compared with the analytical result for the thermodynamic limit obtained with the Bethe ansatz.
	The on-site interaction parameter $U/t = 8$ is chosen as an example of an intermediate coupling regime. In both the real-space and the reciprocal-space basis representation, the wavefunction is highly spread-out; a single Slater determinant is a bad approximation in both cases, thus making it a difficult system to treat. For $U/t = 4$, using a reciprocal-space basis would normally be more suitable as the wavefunction would be more compact in this case. However, we stick to the real-space basis because FCIQMC benefits from the absence of a sign problem significantly more than compactness. For the 102-site systems, we also calculate both with open boundary conditions (OBC) and periodic boundary conditions (PBC). The DMRG computational effort for 1D systems with OBC typically scales as $M^3$, whereas it scales as $M^5$ for PBC. There are, however, similar MPS-based algorithms available that improve the scaling of PBC calculations compared to traditional DMRG~\cite{pippan_mps_2010, dey_dmrg_2016}. In contrast,  the computational cost of FCIQMC is the same for open and periodic boundary conditions as long as the system contains an odd number of spin-up and spin-down electrons such that it remains strictly sign-problem-free. While there is still a small bias for the guided but uncorrected result, the corrected FCIQMC result agrees well with DMRG within statistical error bars. The larger statistical error bars for the $U/t = 4$ compared to the $U/t = 8$ for comparable wall clock time are due to the fact that the real-space basis is less suitable for lower~$U/t$.
	
	Additionally, we also look at the 102-site system with four holes (with total spin projection~$M_s = 0$) at $U/t = 8$. As there are more low-lying excited states, it takes more iterations for FCIQMC to project out the ground state. This results in slightly larger error bars compared to the equivalent half-filled system. Systems off half-filling are also more difficult to treat with DMRG, and thus higher~$M$ values are required to get a converged result. Still, there is good agreement between the DMRG benchmark result and the corrected FCIQMC result.
	
	In general, the FCIQMC results can be continuously improved by adding more CPU time which, is typically the bottleneck to reduce statistical error bars. However, the required memory to store the stochastic representation of the FCIQMC wavefunction at any time during the simulation is typically much lower than the corresponding MPS in DMRG, even though an MPS is already a very efficient way of expressing the ground-state wavefunction of a 1D system. The basic memory requirements of an FCIQMC calculation are the list of the instantaneously occupied determinants in a binary format (with the length of this binary representation depending linearly on the number of sites) and the instantaneous walker population on these determinants. Additionally, there is the need to store the hash table for the main walker list (requiring a hash table with two integers per entry and a list containing empty indices, both roughly scaling with the total number of determinants) and the spawning array (in a conservative estimation requiring up to 1/10 of the main walker list). Thus, in total, one needs to store
	\begin{equation}\label{eq:mem_fciqmc}
	n_\mathrm{int64}^\mathrm{FCIQMC} = \Biggl( \left\lceil\frac{2\ell}{64}\right\rceil + 1 + 3 + \frac{1}{10} \left\lceil\frac{2\ell}{64}\right\rceil \Biggr) \times N_\mathrm{dets}^\mathrm{max}
	\end{equation}
	64-bit integers, where $\ell$ is the number of sites and $N_\mathrm{dets}^\mathrm{max}$ is the maximum number of occupied determinants throughout the simulation. Based on this, we can approximate the memory usage: e.\,g., the periodic 102-site system at $U/t = 8$ was calculated with $3 \times 10^7$~walkers. Since the wavefunction is highly spread out, we can assume that the walkers also occupy roughly $3 \times 10^7$~determinants. According to Eq.~\eqref{eq:mem_fciqmc}, this then requires $2.52 \times 10^8$ 64-bit integers which equals $2.02\,\mathrm{GB}$. This amount is stored as distributed memory as every CPU thread only needs to access data about the determinants that reside on it. However, this is not a fundamental lower limit. In principle, this number could be easily reduced further by running with fewer walkers for more iterations.
	In DMRG, on the other hand, the memory requirement scales quadratically with the bond dimension. The MPS itself requires to store $4\ell$ arrays of size $M \times M$. Another $4 M^2$~doubles is used to store the derivative at a given site during a sweep. To speed up the calculation of the derivative, usually a contraction of $\braket{\psi|\hat{H}|\psi}$ over left and right neighboring sites for each site is also kept in memory, requiring an additional $2 M^2 D_{\hat{H}} \ell$~doubles. $D_{\hat{H}}=4$ is the bond dimension of the 1D Hubbard Hamiltonian when assuming that the h.\,c.~terms are calculated at runtime and are not stored. This sums up to
	\begin{equation}
	n^\mathrm{DMRG}_\mathrm{doubles} = M^2 \bigl[ 4(\ell+1) + 2\ell D_{\hat{H}} \bigr]
	\end{equation}
	double precision floats. For the exemplary 102-site system a bond dimension of $M = 2000$ was used (see table~\ref{table:hubbard_dmrg_bond_dims}). This results in approximately $4.91 \times 10^9$ doubles, requiring $39.3\,\mathrm{GB}$ of memory.
	
	{ 
		\renewcommand{\arraystretch}{1.5}
		\begin{table}[t]
			\begin{tabular}{@{\extracolsep{6pt}}lllll}
				System & Sites & $N/10^6$ & $g$ & $n$\\\hline \hline
				$U/t = 4$, half-filling & $102$ & $50$ & $-0.17$ & $2560$\\
				$U/t = 8$, half-filling & $102$ & $30$ & $-0.15$ & $5120$ \\
				& $150$ & $50$ & $-0.15$ & $5120$\\
				$U/t = 8$, 4 holes ($M_s = 0$) & $102$ & $30$ & $-0.15$ & $2560$\\
				\hline\hline
			\end{tabular}
			\caption{\label{table:hubbard_fciqmc_details}
				Technical details for the FCIQMC calculations of the Hubbard model. Listed are the number of walkers~$N$, the Gutzwiller parameter~$g$ and the number of terms~$n$ in the expansion of the population bias correction. The optimal Gutzwiller is obtained by the projection scheme presented in~\cite{Dobrautz2019}. Optimization of~$g$ using the scheme presented in Fig.~\ref{fig:optimize} only slightly differs.
			}
			
		\end{table}
	}
	
	{ 
		\renewcommand{\arraystretch}{1.5}
		\begin{table}[t]
			\begin{tabular}{@{\extracolsep{6pt}}llll}
				System & Sites & $M$ (PBC) & $M$ (OBC) \\ \hline \hline
				$U/t = 4$, half-filling & $102$ & $2000$ & $300$\\
				$U/t = 8$, half-filling & $102$ & $2000$ & $300$\\
				& $150$ & $4000$ & -- \\
				$U/t = 8$, 4 holes ($M_s = 0$) & $102$ & $5000$ & $1000$\\
				\hline\hline
			\end{tabular}
			\caption{\label{table:hubbard_dmrg_bond_dims}
				Bond dimensions~$M$ for the calculation of the DMRG results in Table~\ref{table:hubbard}.
			}
		\end{table}
	}
	\subsection{Heisenberg Model}\label{sec:heisenberg}
	When the Coulomb repulsion $U$  in the Hubbard model is much larger than the hopping parameter $t$, double occupancies become energetically unfavorable. 
	One can then project the Hubbard Hamiltonian to the space of states with no double occupations and derive an effective Hamiltonian describing the low-energy excitations.
	At half-filling, these excitations are spin excitations, and the effective Hamiltonian is  the spin-$1/2$ antiferromagnetic Heisenberg model with a coupling constant $J=4t^2/U$:
	\begin{equation}\label{eq:heisenberg}
	\hat{H} = J \sum_{\left<ij\right>} \mathbf{S}_i . \mathbf{S}_j\;,
	\end{equation}
	where $\mathbf{S}_i  = (S_i^x, S_i^y, S_i^z)$ is the spin operator on site $i$ with
	\begin{align}
	S_i^x &= \frac{1}{2} \left( \hat{c}^\dagger_{i, \uparrow} \hat{c}_{i, \downarrow} +  \hat{c}^\dagger_{i, \downarrow} \hat{c}_{i, \uparrow}  \right) \\
	S_i^y &= \frac{i}{2} \left(\hat{c}^\dagger_{i, \downarrow} \hat{c}_{i, \uparrow} - \hat{c}^\dagger_{i, \uparrow} \hat{c}_{i, \downarrow}  \right) \\		
	S_i^z &= \frac{1}{2} \left( \hat{c}^\dagger_{i, \uparrow} \hat{c}_{i, \uparrow} - \hat{c}^\dagger_{i, \downarrow} \hat{c}_{i, \downarrow}  \right)	
	\end{align}
	The relation between the two Hamiltonians can be derived using second-order perturbation theory in the $t/U$ parameter, which shows that neighboring electrons with opposite spins can benefit from virtual hopping forming double occupancies.
	The Heisenberg model is a basic and important model for studying magnetic materials and their phase transitions. 
	Like the Hubbard model, it can be solved in the one-dimensional case using the Bethe ansatz~\cite{Karabach1997, Karbach1998}. 
	In two dimensions, the model has been solved numerically to a high-accuracy for different lattice sizes using different methods including Green's function Monte Carlo (GFMC)~\cite{Trivedi1990, Runge1992, Runge1992b}, stochastic series expansion (SSE)~\cite{Sandvik1997}, and world line loop/cluster methods (WLQMC)~\cite{Wiese1994, Evertz2003, alps1, alps2}.
	The model has also been used as a benchmark for recent variational ansatzes such as projected entangled pair states (PEPS)~\cite{Lubasch2014, Liu2017}, neural quantum states (NQS)~\cite{Carleo2017}, and neural autoregressive quantum states (NAQS)~\cite{Sharir2020}.
	
	In FCIQMC, the Heisenberg model is sign-problem-free on a bipartite lattice~\cite{Spencer2012}.
	Walkers of different signs do not mix on such lattices, although all spawns have opposite signs to their parents.
	To see why, we split the lattice into two interlacing sub-lattices, where each site on one sub-lattice is surrounded only by the sites of the other.
	Then we can classify determinants into two groups, $A$ and $B$, depending on whether a determinant has an even or an odd number of spin-up electrons on one of these sub-lattices.
	Applying the Hamiltonian to a determinant increases the number of spin-up electrons on one of the sub-lattices by one and decreases it on the other.  
	Therefore, the Hamiltonian connects members of $A$ only to other members of $B$ and vice versa, so walkers of opposite signs never mix.
	
	In Tables~\ref{table:heisenberg_periodic} and \ref{table:heisenberg_open}, we compare the results of FCIQMC with other methods for different lattice sizes using periodic and open boundary conditions, respectively.
	For periodic boundary conditions (PBC), we used the value $g=0.3$ for the guiding wavefunction and the value $n=2560$ as a correction order, while for open boundary conditions (OBC) we used the values $g=0.35$ and $n=1280$.
	For the lattice sizes $10\times 10$, $16\times 16$ and $32\times 32$, we ran the calculations using $10^5$, $10^7$ and $10^8$ walkers, respectively. 
	The respective numbers of iterations are $16\times 10^6$, $10^6$, and $1.4 \times 10^5$, with an additional $10^5$ iterations of thermalization.
	
	For periodic boundary conditions, we see that FCIQMC results agree extremely well with SSE and is more accurate than GFMC.
	The specific variant of GFMC used for obtaining these results is close in spirit to FCIQMC and formulated similarly as a projective Monte Carlo method in the Hilbert space~\cite{Runge1992b}.
	The two methods differ mainly in the details of their walker dynamics and population control.
	As pointed out in Ref.~\cite{Sandvik1997}, the discrepancy between SSE  (and thus also FCIQMC) results and GFMC results is likely to be due to a remaining population control bias in the latter.
	For open boundary conditions, FCIQMC provides the same level of accuracy as the benchmark results of WLQMC.
	These two QMC methods provide, as expected, much lower energies than the other variational methods: PEPS and NAQS.	
	The results confirm our approach's effectiveness in eliminating the population control bias and demonstrate the potential of FCIQMC  in producing accurate results for large spin systems.
	Nonetheless, we note that the population control bias and the error bars of FCIQMC get larger as the lattice size increases.
	We attribute this behavior to two factors: the shortcomings of our simple guiding wavefunction and the system's closing gap in the thermodynamic limit.

	One obvious drawback of the guiding wavefunction~\eqref{eq:gutzwiller}, in this case, is breaking the rotational symmetry of the Hamiltonian. 
	To see this, remember that for the Heisenberg model, we use the eigenfunctions of the $S_i^z$ operators as  a single-particle basis and thus the diagonal Hamiltonian elements are only a function of the z-component of the spin
	\begin{equation}
	H_{kk} = J \sum_{\left<ij\right>} \braket{D_k| S^z_{i} S^z_{j} |D_k}\;.
	\end{equation}
	However, the Heisenberg Hamiltonian of Eq.~\eqref{eq:heisenberg} is isotropic and thus the guiding wavefunction does not possess the necessary rotational invariance  of the true ground state.
	The discrepancy between the guiding wavefunction and the exact ground state evidently becomes more critical as the systems size increases. It thus lessens the effectiveness of importance sampling in reducing the population control bias and the statistical error bars of the guided results.
	Moreover, it is has been shown that the gap between the ground state and the first excited state of the Heisenberg model scales inversely with the lattice size~\cite{Neuberger89, Gross1989}, which implies longer projection times.
	As a result, more correction terms need to be included in the reweighting procedure to remove the same amount of bias, which increases the statistical error bars of the corrected results further.	
	While the closing gap is an inherent property of the system under study, the guiding wavefunction is under our control and can be improved to refine the results further and accelerate convergence.
	A promising research direction would be employing one of the variational anastzes as a guiding wavefunction in FCIQMC.
	Particularly interesting are neural quantum states that provide compact representation and allow fast evaluations of the guiding wavefunction, a necessary property for its practical use in FCIQMC.
	We would like to mention that a similar increase in the computational effort required to overcome the population control bias has been observed in recent applications of DMC to other spin systems.
	It was found that using more flexible guiding wavefunction helps 
	
	{ 
		\renewcommand{\arraystretch}{1.5}
		\begin{table}[t]
			\begin{tabular}{@{\extracolsep{6pt}}l|l|l}
				Lattice & Method       & Energy  per Site \\ \hline \hline
				\multirow{4}{*} { $10\times 10$ }			
				& GFMC &$-0.671~492(27)$ \\			 			
				& SSE     & $-0.671~549(4)$ \\ 
				\cline{2-3}
				& FCIQMC -- Guided           &$-0.671~541(1)$ \\
				& FCIQMC -- Corrected &$-0.671~553(1)$ \\			
				\hline\hline
				
				\multirow{4}{*} { $16\times 16$ }
				& GFMC &$-0.669~872(28)$ \\			 			
				& SSE     & $-0.669~976(7)$ \\
				\cline{2-3} 
				& FCIQMC -- Guided           &$-0.669~960(5)$ \\
				& FCIQMC -- Corrected &$-0.669~976(8)$ \\
				\hline\hline
				
				\multirow{3}{*} { $32\times 32$ }
				& SSE     & $-0.669~5115(8)$ \\
				\cline{2-3} 
				& FCIQMC -- Guided           & $-0.669~487(9)$ \\
				& FCIQMC -- Corrected & $-0.669~492(10)$ \\
				\hline\hline
				
			\end{tabular}
			\caption{\label{table:heisenberg_periodic}
				Results for the Heisenberg model using periodic boundary conditions.
				GFMC results are taken from Ref.~\cite{Runge1992b}, while SSE results are taken from Refs.~\cite{Sandvik1997,Sandvik2010}.			
			}
			
		\end{table}
	}
	
	{ 
		\renewcommand{\arraystretch}{1.5}
		\begin{table}[t]
			\begin{tabular}{@{\extracolsep{6pt}}l|l|l}
				Lattice & Method       & Energy per Site  \\ \hline \hline
				
				\multirow{5}{*} { $10\times 10$ }
				& PEPS    & $-0.628~601(2)$ \\ 			
				& NAQS  &$-0.628~627(1)$ \\			 			
				& WLQMC   & $-0.628~656(2)$ \\
				\cline{2-3}
				& FCIQMC -- Guided           &$-0.628~646(1)$ \\
				& FCIQMC -- Corrected &$-0.628~656(1)$ \\			
				\hline\hline
				
				\multirow{5}{*} { $16\times 16$ }
				& PEPS   &  $-0.643~391(3)$\\ 			
				& NAQS &$-0.643~448(1)$ \\
				& WLQMC   & $-0.643~531(2)$ \\ 
				\cline{2-3}
				& FCIQMC -- Guided           &$-0.643~530(2)$ \\
				& FCIQMC -- Corrected &$-0.643~531(3)$ \\			
				\hline\hline			
			\end{tabular}
			\caption{\label{table:heisenberg_open}
				Results for the Heisenberg model using open boundary conditions.
				WLQMC, PEPS and NAQS results are obtained respectively from Refs.~\cite{Lubasch2014, Liu2017, Sharir2020}.
			}
		\end{table}
	}
	
	\section{Summary}
	In this paper, we related the population control bias of FCIQMC directly to the covariance between the sampled wavefunction and the shift.
	We used this relation to explain why the bias scales inversely with the average number of walkers and quantify the resulting discrepancy between the different energy estimators.
	We then derived a post-processing reweighting procedure that practically eliminates the bias from different energy estimates, including the average shift.
	We also showed how importance sampling could dramatically reduce the bias and reduce the required number of correction terms and the associated error bars.
	Both techniques have only a small computational cost and make FCIQMC results practically exact (within statistical error bars) in the absence of a sign problem.
	Our approach's effectiveness makes FCIQMC competitive with other high-accuracy methods in its application to large sign-problem-free systems.
	Yet, its value is not restricted to these systems but extends to a broader range of systems with sign problems, whose study is the topic of an upcoming publication.
	
	\section*{Acknowledgments}
	A.A. acknowledges discussions with Joachim Brand on population control bias. K.G. thanks Erik Koch for helpful discussions about the Gutzwiller wavefunction. The authors gratefully acknowledge funding from the Max Planck Society.
	
	\section*{Note Added}
	A recent preprint by \textcite{brand2021}, which also studies the population control bias of FCIQMC, has been announced since the initial submission of this manuscript. 
	Their work is largely complementary to ours and provides further insight into understanding the biased dynamics of FCIQMC in terms of It\^{o} stochastic differential equations.
	They also present results for the Bose-Hubbard model and demonstrate numerically that the bias of this system scales as a power-law with respect to the total number of walkers rather than linearly \footnote{
		While we have not observed such a power-law scaling in our calculations, our analysis in Sec.~\ref{sec:scaling} does not rule this behavior out.
		In Eq.~\ref{eq:cov}, we use Taylor's approximation of the logarithm of the total number of walkers, which can break down if the samples deviate too much from their mean value.
		Furthermore, the authors of Ref.~\cite{brand2021} point out that they are using the integer-walker formulation of FCIQMC, in contrast to the real-walker formulation used in our work, and that this may be contributing to the observed power-law behavior.}. They do not, however, apply importance sampling, which is expected to ameliorate the bias substantially.
	
	\appendix
	\section{Periodic Shift Update}\label{app:delay}
	When the shift is updated every $A$ iterations instead of continuously, we
	need to change the invariant quantity of Eq.~\eqref{eq:invariant} to
	\begin{equation}
	S(\tau) + \frac{\gamma}{A \Delta \tau} \log N(\tau^\prime)\;,
	\end{equation}
	where $\tau^\prime= A \times (\tau \bmod A)$ is the last time the shift got updated. 
	Then the results using a continuous update still hold if we redefine the relative bias terms~$B_i$ using quantities at the appropriate time steps
	\begin{align}
	B_i(\tau) &\coloneqq -\operatorname{Cov}\left[N_i(\tau), S(\tau)\right] \frac{\overline{N(\tau^\prime)}}{\overline{N_i(\tau)}} \nonumber \\
	&\approx   \frac{\gamma}{A \Delta \tau } \frac{\operatorname{Cov}\left[N_i(\tau), N(\tau^\prime)\right]}{\overline{N_i(\tau)}}
	\;.
	\end{align}
	Although $N_i$ and $N$ above correspond to different iterations, their covariance still scales linearly with the number of walker as in Eq.~\eqref{eq:rel_scaling}.
	The main difference is that we expect this covariance to be weaker because the correlation between quantities at different iterations gets smaller when they are further apart. 
	
	\section{Covariance of the Shift and the Number of Walkers}\label{app:cov}
	The covariance between the shift and the number of walkers can be related to the variance in the number of walkers as following:
	\begin{equation}
	\operatorname{Cov}\left(N, S\right)  = \sum_{i}\operatorname{Cov}\left(N_i, S\right) \approx -\frac{\gamma}{A\Delta\tau} \frac{\operatorname{Var}\left(N\right)}{\overline{N}}\;,
	\end{equation}
	where Eq.~\eqref{eq:cov} is used as an approximation.
	It is important to note here that $\operatorname{Cov}\left(N, S\right) $ does not scale with parameters $\gamma, A, \Delta \tau$ and $\overline{N}$ as suggested by the above formula.
	The reason is that $\operatorname{Var}\left(N\right) $ implicitly depends on those parameters and scales in the opposite direction.
	For example, by using half the value of $\gamma$, the variance of the number of walkers will almost double due to the looser control of the shift.
	The behavior becomes more evident by rewriting the covariance in terms of the shift variance
	\begin{equation}
	\operatorname{Cov}\left(N, S\right)\approx -\frac{A\Delta\tau }{\gamma} \operatorname{Var}\left(S\right) \overline{N}\;,
	\end{equation}
	which leads to the exact opposite explicit dependence on these parameters.
	This argument can be extended to the covariance between the shift and the individual populations $N_i$ and explains why we could not establish a clear relation between the population control bias and the parameters $\gamma, A$ or $\Delta \tau$.
	
	%
	
\end{document}